\documentclass[12pt]{article}

\usepackage{amssymb}
\usepackage{graphicx}
\usepackage{enumerate}

\newcommand{\sect}[1]{\section{#1}\setcounter{equation}{0}}

\begin{document}
\baselineskip 16pt

\def\ad{a^\dagger}

\def\chibar{\bar{\chi}}
\def\G{\Gamma}
\def\L{\Lambda}
\def\oG{|\G|}
\def\Mg{M/\G}
\def\s3g{S^3/\G}
\def\fphi{\varphi}
\def\gsim{\; \raisebox{-.8ex}{$\stackrel{\textstyle >}{\sim}$}\;}
\def\lsim{\; \raisebox{-.8ex}{$\stackrel{\textstyle <}{\sim}$}\;}
\def\beq{\begin{equation}}
\def\eeq{\end{equation}}
\def\bq{\begin{quote}}
\def\eq{\end{quote}}
\def\bea{\begin{eqnarray}}
\def\eea{\end{eqnarray}}
\def\ben{\begin{enumerate}}
\def\een{\end{enumerate}}
\def\bit{\begin{itemize}}
\def\eit{\end{itemize}}
\def\({\left (}
\def\){\right )}
\def\[{\left [}
\def\]{\right ]}
\def\T{{\bf T}}

\def\fphi{\varphi}

\def\g{\gamma}
\def\a{\alpha}
\def\b{\beta}
\def\d{\delta}
\def\k{\kappa}
\def\l{\lambda}
\def\m{\mu}
\def\n{\nu}
\def\r{\rho}
\def\s{\sigma}
\def\o{\omega}
\def\O{\Omega}
\def\S{\Sigma}
\def\t{\tau}
\def\z{\zeta}
\def\D{\Delta }
\def\tphi{\tilde{\fphi}}
\def\ra{\rangle}
\def\la{\langle}
\def\p{\partial}
\def\bx{{\bf x}}
\def\bk{{\bf k}}
\def\bl{{\bf l}}
\def\kdx{\bk\cdot\bx}

\def\R{{\cal R}}
\def\W{{\cal W}}
\def\V{{\cal V}}

\def\half{\textstyle{\frac{1}{2}}}
\bigskip
\hspace*{\fill}
\bigskip\bigskip\bigskip

\begin{center}
\Large \bf Introduction to Quantum Fields in Curved Spacetime
and the Hawking Effect
\end{center}
\bigskip\bigskip\bigskip
\centerline{\large Ted Jacobson}
\medskip
\centerline{\it Department of Physics, University of Maryland}
\centerline{\it College Park, MD 20742-4111}
\centerline{\tt jacobson@physics.umd.edu}

\begin{abstract}
These notes  introduce
the subject of quantum field theory in curved spacetime
and some of its applications
and the questions they raise. Topics include particle creation
in time-dependent metrics, quantum origin of primordial
perturbations, Hawking effect, the trans-Planckian question, 
and Hawking radiation on a lattice.
\vskip 1cm
\begin{center}
{\it Based on 
lectures given at the  CECS School on Quantum Gravity 
in Valdivia, Chile, January 2002.}
\end{center}
\end{abstract}

\newpage

\tableofcontents

\newpage

\sect{Introduction}
\label{Intro}

Quantum gravity remains an outstanding problem of fundamental
physics. The bottom line is we don't even know the nature
of the system that should be quantized. The spacetime
metric may well be just a collective description of some more basic
stuff. The fact\cite{tos} that the semi-classical Einstein equation can be
derived by demanding that the first law of thermodynamics
hold for local causal horizons, assuming the proportionality
of entropy and area, leads one to suspect that the metric is
only meaningful in the thermodynamic limit of something else.
This led me at first to suggest that the metric shouldn't be
quantized at all. However I think this is wrong. Condensed matter
physics abounds with examples of collective modes that become
meaningless at short length scales, and which are nevertheless
accurately treated as quantum fields within the appropriate domain.
(Consider for example the sound field in a Bose-Einstein condensate
of atoms, which loses meaning at scales below the so-called
``healing length",
which is still several orders of magnitude longer than the atomic
size of the fundamental constituents.)
Similarly,
there exists a perfectly good perturbative approach to quantum
gravity in the framework of low energy effective field theory\cite{Donoghue:1997hx}.
However, this is not regarded as a solution to the problem of quantum
gravity, since the most pressing questions are non-perturbative in nature:
the nature and fate of spacetime singularities, the fate of Cauchy horizons,
the nature of the microstates counted by black hole entropy, 
and the possible
unification of gravity with other interactions.

At a shallower level,
the perturbative approach of effective field theory
is nevertheless relevant both for its indications about
the deeper questions and for its application to physics
phenomena in their own right. It
leads in particular to the subject of quantum
field theory in curved spacetime backgrounds,
and the ``back-reaction" of the quantum fields on such
backgrounds. Some of the most prominent of these
applications are the Hawking radiation by black holes,
primordial density perturbations, and early universe phase transitions.
It also fits into the larger category of quantum field theory (qft) 
in inhomogeneous
and/or time-dependent backgrounds of other fields or
matter media, and is also intimately tied to non-inertial
effects in flat space qft such as the Unruh effect.
The pertubative approach is also known as ``semi-classical
quantum gravity", which refers to the setting where there
is a well-defined classical background geometry about which
the quantum fluctuations are occuring.

The present notes are an introduction
to some of the essentials and phenomena 
of quantum field theory in curved spacetime. Familiarity with 
quantum mechanics and general relativity are assumed.
Where computational steps are omitted I expect that the
reader can fill these in as an exercise. 

Given the importance of the subject, it is curious that
there are not very many books dedicated to it.
The standard reference by Birrell and Davies\cite{BD} was published twenty years 
ago, and another monograph by Grib, Mamaev, and Mostapanenko\cite{Grib}, 
half of which addresses strong background field effects in flat spacetime, 
was published two years earlier originally in Russian and then
in English  ten years ago. Two books with a somewhat more limited scope
focusing on fundamentals with a mathematically rigorous
point of view are those by Fulling\cite{Fulling89} and Wald\cite{Wald94}.
This year DeWitt\cite{brycebook} 
published a comprehensive two volume treatise with a much wider scope
but including much material on quantum fields in curved spacetime.
A number of review articles (see e.g. \cite{DeWitt75,Isham, Gibbons, DeWittEC,
FullRuij, Brout:rd}) 
and many shorter introductory lecture notes (see e.g. \cite{Ford,Wipf,Traschen})
are also available. For more information on topics not explicitly
referenced in the text of these notes the above references should be consulted.

In these notes the units are chosen with $c=1$ but $\hbar$ and $G$ are kept 
explicit. The spacetime signature is $({+}{-}{-}{-})$. Please send
corrections if you find things here that are wrong.

\sect{Planck length and black hole thermodynamics}

Thanks to a scale separation 
it is useful to distinguish quantum field theory 
in a curved background spacetime (qftcs)
from  true quantum gravity (qg).
Before launching into the qftcs formalism, 
it seems worthwhile to have a quick look
at some of the interesting 
issues that qftcs is  concerned with.

\subsection{Planck length}
It is usually presumed that the length scale of quantum gravity is the
Planck length $L_P = (\hbar G/c^3)^{1/2}\approx 10^{-33}$ cm.
The corresponding energy scale is $10^{19}$ GeV.
Recent  ``braneworld scenarios", in which 
our 4d world is a hypersurface in a higher dimensional
spacetime,  put the scale of quantum gravity much
lower, at around a TeV, corresponding to 
$L_{\rm TeV}=10^{16}L_P\approx 10^{-17}$ cm.
In either case, there is plenty of room
for applicability of qftcs. (On the other hand, we are much closer
to seeing true qg effects in TeV scale qg. For example we might
see black hole creation and evaporation in cosmic rays or accelerators.)

Here I will assume Planck scale qg, and look at some dimensional
analysis to give a feel for the phenomena.
First, how should we think of the Planck scale? The
Hilbert-Einstein action is 
$S_{HE}=(\hbar/16\pi L_P^{2})\int d^4x |g|^{1/2}R$.
For a spacetime region with radius of 
curvature $L$ and 4-volume $L^4$  
the action is $\sim\hbar (L/L_P)^2$. This suggests that
quantum curvature fluctuations with radius 
less than the Planck length $L\lsim L_P$ 
are unsuppressed. 

Another way to view the significance of the Planck length is 
as the minimum localization length $\D x$, in the sense
that if $\D x<L_P$ a black hole swallows the $\D x$.
To see this, note that the uncertainty relation 
$\D x\D p\ge\hbar/2$ implies
$\D p\gsim \hbar/\D x$ which implies $\D E\gsim \hbar c/\D x$.
Associated with this uncertain energy is a Schwarzschild radius
$R_s(\D x)=2G\D M/c^2=2G\D E/c^4$,
hence quantum mechanics and gravity imply 
$R_s(\D x)\gsim L_P^2/\D x$. The uncertain $R_s(\D x)$ is 
less than $\D x$ only if $\D x\gsim L_P$.

\subsection{Hawking effect}
\label{hawkintro}

Before Hawking, spherically symmetric, static 
black holes were assumed
to be completely inert. In fact, it seems more natural that
they can decay, since there is no conservation law preventing
that. The decay is quantum mechanical, and thermal:  Hawking
found that a black hole radiates at a temperature
proportional to $\hbar$, $T_H=(\hbar/2\pi)\k$, where $\k$ is the
surface gravity. The fact that the radiation is thermal is
even natural, for {\it what else could it be}? The very nature of the
horizon is a causal barrier to information, and yet the Hawking
radiation emerges from just outside the horizon. Hence there can be 
no information in the Hawking radiation, save for the mass of the
black hole which is visible on the outside, so it must be a maximum
entropy state, i.e. a thermal state with a temperature determined by the 
black hole mass. 

For a Schwarzschild
black hole $\k=1/4GM=1/2R_s$, so the Hawking temperature is 
inversely proportional to the mass. 
This implies a thermal wavelength
$\l_H=8\pi^2 R_s$, a purely geometrical relationship
which indicates two things. First, although the emission
process involves quantum mechanics, its ``kinematics"
is somehow classical. Second,  as long as the Schwarzschild
radius is much longer than the Planck length,  it should
be possible to understand the Hawking effect using
only qftcs, i.e. semi-classical qg. 
(Actually one must also require that the back-reaction
is small, in the sense that the change in the Hawking temperature
due to the emission of a single Hawking quantum is small. 
This fails to hold for for a very nearly extremal black hole\cite{Preskill:1991tb}.)

A Planck mass ($\sim 10^{-5}$ gm) black 
hole---if it could be treated semi-classically---would have a Schwarzschild
radius of order the Planck length ($\sim 10^{-33}$ cm)
and a Hawking temperature of order the Planck energy ($\sim 10^{19}$ GeV). 
From this the Hawking temperatures for other black holes can be found by scaling.
A solar mass black hole has a Schwarzschild radius
$R_s\sim 3$ km hence a Hawking temperature $\sim10^{-38}$ times
smaller than the Planck energy, i.e. $10^{-19}$ GeV. Evaluated more carefully
it works out to $T_H\sim 10^{-7}$ K. For
a mini black hole of mass $M=10^{15}$ gm one has 
$R_s\sim 10^{-13}$ cm and $T_H\sim 10^{11} {\rm K}\sim 10$ MeV.

The ``back reaction", i.e. the response of the spacetime
metric to the Hawking process, should be well approximated 
by the semi-classical Einstein equation
$G_{\m\n}=8\pi G \la T_{\m\n}\ra$ provided it is a small
effect. To assess the size, we can compare 
the stress tensor to the 
background curvature near the horizon (not
to $G_{\m\n}$, since that vanishes in the background). 
The background Riemann tensor components are $\sim 1/R_s^2$
(in suitable freely falling reference frame), while
for example the energy density is
$\sim T_H^4/\hbar^3\sim \hbar/R_s^4$.
Hence $G \la T_{\m\n}\ra\sim \hbar G/R_s^4=(L_P/R_s)^2 R_s^{-2}$, which is much less than the background curvature provided $R_s\gg L_P$, i.e. provided the black hole is large compared to the Planck length.

Although a tiny effect for astrophysical black holes, 
the Hawking process 
has a profound implication: black holes can ``evaporate". 
How long does one take to evaporate? 
It emits roughly one Hawking quantum per light
crossing time, hence
$dM/dt\sim
T_H/R_S\sim \hbar/R_S^2\sim \hbar/G^2M^2$. In Planck units
$\hbar=c=G=1$, we have $dM/dt\sim M^{-2}$. Integration yields the
lifetime $\sim M^3\sim (R_s/L_P)^2 R_S$. For the $10^{15}$ gm,
$10^{-13}$ cm black hole mentioned earlier we  have $(10^{20})^2 10^{-13} {\rm
cm}=10^{27}$ cm,  which is the age of the universe. Hence a black
hole of that mass in the early universe would be explosively ending its
life now. These are called ``primordial black holes" (pbh's). 
None have been knowingly 
observed so far, nor their detritus, which puts limits on the
present density of pbh's (see for example the references in\cite{Barrau:1999sk}). 
However, it has been suggested\cite{Barrau:1999sk} that they
might nevertheless be the source of the highest energy 
($\sim 3\times 10^{20}$ eV) cosmic rays. 
Note that even if pbh's were copiously
produced in the early universe, their initial number 
density could easily have been inflated away if they formed before inflation.

\subsection{Black hole entropy}

How about the black hole entropy? The thermodynamic relation
$dM=T_H dS=(\hbar/8\pi GM) dS$ (for a black hole with 
no angular momentum or charge) implies
$S_{BH}=4\pi G M^2/\hbar= A_H/4L_P^2$,
where $A_H=4\pi R_S^2$ is the black hole horizon area. 
What is this huge entropy? What microstates does it count?

Whatever the microstates may be, $S_{BH}$ is the
lower bound for the entropy emitted to the outside world 
as the black hole evaporates, i.e. the minimal missing information
brought about by the presence of the black hole. 
By sending energy into a black hole one could make it last
forever, emitting an arbitrarily large amount of entropy, so 
it only makes sense to talk about the minimal entropy.
This lower bound is attained 
when the black hole evaporates reversibly into a thermal bath at
a temperature infinitesimally below $T_H$.
(Evaporation into vacuum is irreversible, so the 
total entropy of the outside increases even more\cite{Zurek},
$S_{\rm emitted}\sim (4/3) S_{BH}$.)
This amounts to a huge entropy increase.
The only way out of this conclusion would be 
if the  semi-classical analysis breaks down
for some reason...which is suspected by many,
not including me.

The so-called ``information paradox" refers to the
loss of information associated with this entropy increase
when a black hole evaporates,
as well as to the loss of any other
information that falls into the black hole. I consider it no
paradox at all, but many think it is a problem to be
avoided at all costs. I think this viewpoint 
results from
missing the strongly non-perturbative role
of quantum gravity in evolving the spacetime 
into and beyond the classically singular region,
whether by generating baby universes or otherwise
(see section \ref{infoloss} for more discussion of this point).
Unfortunately it
appears unlikely that semi-classical qg can ``prove"
that information is or is not lost, though people have
tried hard. For it not
to be lost there would have to be subtle
effects not captured by semi-classical qg, yet
in a regime where semi-classical description 
``should" be the whole story at leading order.

\sect{Harmonic oscillator}

One can study a lot of interesting issues with free fields in
curved spacetime, and I will restrict entirely to that
case except for a brief discussion. 
Free fields in curved spacetime are similar 
to collections of harmonic oscillators with time-dependent
frequencies. I will therefore begin by developing the properties 
of a one-dimensional quantum 
harmonic oscillator in a formalism  parallel to that used
in quantum field theory. 

The action for a particle of mass $m$ moving in a time dependent
potential $V(x,t)$ in one dimension takes the form
\beq S=\int dt\; L\qquad\qquad\ L=\frac{1}{2}m\dot{x}^2 - V(x,t)
\eeq
from which follows the equation of motion $m\ddot{x}=-\partial_x
V(x,t)$. Canonical quantization proceeds by (i) defining the
momentum conjugate to $x$, $p=\partial L/\partial
\dot{x}=m\dot{x}$, (ii) replacing $x$ and $p$ by operators
$\hat{x}$ and $\hat{p}$, and (iii) imposing the canonical
commutation relations $[\hat{x},\hat{p}]=i\hbar$. The operators
are represented as hermitian linear operators on a Hilbert space,
the hermiticity ensuring that their spectrum is real as befits a
quantity whose classical correspondent is a real number.\footnote{In 
a more abstract algebraic approach, one does not
require at this stage a representation but rather requires that
the quantum variables are elements of an algebra equipped
with a star operation satisfying certain axioms. For quantum
mechanics the algebraic approach is no different from the
concrete representation approach, however for quantum fields
the more general algebraic approach turns out to be necessary
to have sufficient generality. In these lectures I will ignore
this distinction. For an introduction to the algebraic approach
in the context of quantum fields in curved spacetime see
\cite{Wald94}. For a comprehensive treatment of the
algebraic approach to quantum field theory see
\cite{Haag}.\label{algebra}}
In the Schr\"odinger picture the state is time-dependent, and the
operators are time-independent. In the position representation
for example the momentum operator is given by
$\hat{p}=-i\hbar\partial_x$. In the Heisenberg picture the state
is time-independent while the operators are time-dependent. The
commutation relation then should hold at each time, but this is
still really only one commutation relation since the the equation
of motion implies that if it holds at one initial time it will
hold at all times. In terms of the position and velocity, the
commutation relation(s) in the Heisenberg picture take the form
\beq [x(t),\dot{x}(t)]=i\hbar/m. \label{cr}\eeq
Here and from here on the hats distinguishing numbers from
operators are dropped.

Specializing now to a harmonic oscillator potential $V(x,t)=\frac{1}{2}
m\o^2(t)x^2$ the equation of motion takes the form
\beq \ddot{x}+\o^2(t)x=0.\eeq
\label{ho(t)}
Consider now any operator solution $x(t)$ to this equation. Since
the equation is second order the solution is determined by the two
hermitian operators $x(0)$ and $\dot{x}(0)$, and since the
equation is linear the solution is linear in these operators. It
is convenient to trade the pair $x(0)$ and $\dot{x}(0)$ for a
single time-independent
non-hermitian operator $a$, in terms of which the solution
is written as
\beq x(t) = f(t) a + \bar{f}(t) \ad,\label{xoperator}\eeq
where $f(t)$ is a complex function satisfying the classical
equation of motion,
\beq \ddot{f}+\o^2(t)f=0,\label{fho}\eeq
$\bar{f}$ is the complex conjugate of $f$, and $\ad$ is the
hermitian conjugate of $a$. The commutation relations (\ref{cr})
take the form
\beq \la f,f\ra\, [a,\ad]=1, \label{cr2}\eeq
 where the
bracket notation is defined by
\beq \la f,g\ra=(im/\hbar)\Bigl(\bar{f}\p_t{g}-(\p_t{\bar{f}})g\Bigr).
\label{bracket}\eeq
If the functions $f$ and $g$ are solutions to the harmonic
oscillator equation (\ref{fho}), then the bracket
(\ref{bracket}) is independent of the time $t$ at which the
right hand side is evaluated, which is consistent with the 
assumed time independence of $a$.

Let us now assume that the solution $f$ is chosen so that the 
real number $\la f,f\ra$ is positive. Then
by rescaling $f$ we can arrange to have 
\beq
\la f,f\ra=1.
\label{fnorm}
\eeq
In this case the commutation relation (\ref{cr2}) becomes
\beq [a,\ad]=1, \label{cra}\eeq
the standard relation
for the harmonic oscillator raising and lowering operators.  
Using the bracket with the operator $x$ we can pluck out the
raising and lowering operators from the position operator,
\beq a = \la f,x\ra,\qquad\qquad \ad=-\la\bar{f},x\ra.
\label{pluckout}\eeq
Since both $f$ and $x$ satisfy the equation of motion, the
brackets in (\ref{pluckout}) are time independent as they must
be.

A Hilbert space representation of the operators 
can be built by introducing a state $|0\ra$
defined to be normalized and satisfying $a|0\ra=0$. 
For each $n$, the state
$|n\ra=(1/\sqrt{n!})(\ad)^n|0\ra$ is a normalized
eigenstate of the number operator $N=\ad a$ with
eigenvalue $n$. The span of all these states defines a Hilbert
space of ``excitations" above the
state $|0\ra$.

So far the solution $f(t)$ is arbitrary, except for the normalization
condition (\ref{fnorm}). A change in $f(t)$ could be accompanied
by a change in $a$ that keeps the solution $x(t)$ unchanged.
In the special case of a constant frequency 
$\o(t)=\o$ however, the energy is conserved, and a special choice of $f(t)$
is selected if we require that the state $|0\ra$ be the
ground state of the Hamiltonian. Let us see how this comes about.

For a general $f$ we have
\bea
H&=&\half m \dot{x}^2 + \half m\o^2 x^2\\
&=&\half m\left[(\dot{f}^2 +\o^2f^2) aa+(\dot{f}^2 +\o^2f^2)^*\ad \ad
+(|\dot{f}|^2 +\o^2|f|^2)(a\ad + \ad a)\right]\!.
\label{ham1}
\eea
Thus
\beq
H|0\ra=\half m(\dot{f}^2 +\o^2f^2)^*\ad \ad|0\ra
+(|\dot{f}|^2 +\o^2|f|^2)|0\ra,
\label{Hgnd}
\eeq
where the commutation relation (\ref{cra}) was used in the 
last term. If $|0\ra$ is to be an eigenstate of $H$, 
the first term must vanish, which requires
\beq
\dot{f}=\pm i\o f. 
\label{posnegfreq}
\eeq
For such an $f$ the norm is
\beq
\la f,f\ra = \mp \frac{2m\o}{\hbar}|f|^2,
\eeq
hence the positivity of the 
normalization condition (\ref{fnorm})
selects from (\ref{posnegfreq}) the 
minus sign. This yields what is called 
the normalized {\it positive frequency}
solution to the equation of motion,
defined by 
\beq
f(t)=\sqrt{\frac{\hbar}{2m\o}}\, e^{-i\o t}
\label{posfreq}
\eeq
up to an arbitrary constant phase factor.  

With $f$ given by (\ref{posfreq})
the Hamiltonian 
(\ref{ham1})  becomes
\bea
H&=&\half \hbar\o (a\ad + \ad a)\\
&=&\hbar\o(N + \half),
\label{ham2}
\eea
where the commutation relation (\ref{cra}) was used in the 
last step. The spectrum of the number operator 
is the non-negative integers, hence the minimum energy state
is the one with $N=0$, and ``zero-point energy" $\hbar \o/2$. 
This is just the state $|0\ra$ 
annihilated by $a$ as defined above.  
If any function other than (\ref{posfreq})
is chosen to expand the
position operator as in (\ref{xoperator}), the state
annihilated by $a$ is not the ground state
of the oscillator. 

Note that although the mean value 
of the position is zero in the ground state,
the mean of its square is  
\beq
\la0|x^2|0\ra=\hbar/2m\o.
\label{zpf}
\eeq
This characterizes the ``zero-point fluctuations" of the position
in the ground state.

\sect{Quantum scalar field in curved spacetime}
Much of interest can be done with a scalar field, so it suffices
for an introduction. The basic concepts and methods extend
straightforwardly to tensor and spinor fields. To being with let's
take a spacetime of arbitrary dimension $D$, with a metric
$g_{\m\n}$ of signature $(+-\cdots-)$. The action for the scalar
field $\fphi$ is
\beq S=\int d^Dx\;
\sqrt{|g|}\frac{1}{2}\(g^{\m\n}\p_\m\fphi\p_\n\fphi -(m^2+\xi
R)\fphi^2\), \label{scalaraction}\eeq
for which the equation of motion is 
\beq \(\Box + m^2+\xi R\)\fphi=0,\qquad\qquad
\Box=|g|^{-1/2}\p_\m|g|^{1/2}g^{\m\n}\p_\n.
\label{scalareom}\eeq
(With $\hbar$ explicit, the mass $m$ should 
be replaced by $m/\hbar$, however we'll leave 
$\hbar$ implicit here.)
The case where the coupling $\xi$ to the Ricci scalar $R$ vanishes 
is referred to as ``minimal coupling", and that 
equation is called the {\it Klein-Gordon} (KG) equation. 
If also the mass $m$ 
vanishes it is called the ``massless, minimally coupled scalar".
Another special case of interest
is ``conformal coupling" with $m=0$ and $\xi=(D-2)/4(D-1)$.

\subsection{Conformal coupling}
Let me pause briefly to explain the meaning of conformal coupling
since it comes up often in discussions of quantum fields in
curved spacetime, primarily either because Robertson-Walker metrics are
conformally flat or because all two-dimensional metrics are conformally flat.  Consider making a position dependent conformal
transformation of the metric:
\beq \widetilde{g}_{\m\n}=\Omega^2(x){g}_{\m\n},
\label{metricchange}\eeq
which induces the changes
\beq \widetilde{g}^{\m\n}=\Omega^{-2}(x){g}^{\m\n},\qquad
|\widetilde{g}|^{1/2}=\Omega^D(x)|{g}|^{1/2},\qquad
|\widetilde{g}|^{1/2} \widetilde{g}^{\m\n} =\Omega^{D-2}(x)
|{g}|^{1/2}{g}^{\m\n}, 
\eeq
\beq \widetilde{R}=\widetilde{g}^{\m\n}\widetilde{R}_{\m\n}=\Omega^{-2}\Bigl({R}-
2(D-1)\Box\ln\Omega-
(D-1)(D-2)g^{\a\b}(\ln \Omega)_{,\a}(\ln\Omega)_{,\b}\Bigr). \eeq
In $D=2$ dimensions, the action is simply invariant in the
massless, minimally coupled case without any
change of the scalar field: $S[\fphi,g]=S[\fphi,\widetilde{g}]$.
In any other dimension, the kinetic term is invariant under a
conformal transformation with a {\it constant} $\Omega$ if we
accompany the metric change (\ref{metricchange}) with a change
of the scalar field, $\widetilde{\fphi} = \Omega^{(2-D)/2}{\fphi}$.
(This scaling relation corresponds to the fact that the scalar
field has dimension $[{\rm length}]^{(2-D)/2}$ since the action
must be dimensionless after factoring out an overall $\hbar$.)
For a non-constant $\Omega$ the derivatives in the kinetic term
ruin the invariance in general. However it can be shown that
the action is invariant
(up to a boundary term) if the coupling constant $\xi$ is chosen
to have the special value given in the previous paragraph, i.e.
$S[\fphi,g]=S[\widetilde{\fphi},\widetilde{g}]$. In $D=4$
dimensions that value is $\xi=1/6$.

\subsection{Canonical quantization}
To canonically quantize we first pass to the Hamiltonian
description. Separating out a time coordinate $x^0$,
$x^\m=(x^0,x^i)$, we can write the action as
\beq S=\int dx^0\; L,\qquad\qquad L=\int d^{D-1}x\; {\cal L}. \eeq
The canonical momentum at a time $x^0$ is given by
\beq \pi(\underline{x}) =\frac{\d
L}{\d\Bigl(\p_0\fphi(\underline{x})\Bigr)}=|g|^{1/2}g^{\m
0}\p_\m\fphi(\underline{x})=
|h|^{1/2}n^{\m}\p_\m\fphi(\underline{x}).\label{momentum} \eeq
Here $\underline{x}$ labels a point on a surface of constant
$x^0$, the $x^0$ argument of $\fphi$ is suppressed, $n^\m$ is the
unit normal to the surface, and $h$ is the determinant of 
the induced spatial metric $h_{ij}$. To quantize, the
field $\fphi$ and its conjugate momentum $\fphi$ are now promoted
to hermitian operators\footnote{See footnote \ref{algebra}.}
and required to satisfy the canonical commutation relation,
\beq [\fphi(\underline{x}),
\pi(\underline{y})]=i\hbar\d^{D-1}(\underline{x}, \underline{y})
\label{ccr} \eeq
It is worth noting that, being a variational derivative, the
conjugate momentum is a density of weight one, and hence the
Dirac delta function on the right hand side of (\ref{ccr}) is
a density of weight one in the second argument. It is 
defined by the property 
$\int d^{D-1}y\; \d^{D-1}(\underline{x}, \underline{y}) f(\underline{y})=f(\underline{x})$
for any scalar function $f$, without the use of a metric volume element.

In analogy with the bracket (\ref{bracket}) defined for the case
of the harmonic oscillator, one can form a conserved bracket
from two complex solutions to the scalar wave equation
(\ref{scalareom}),
\beq \la f,g\ra=\int_\S d\S_\m\; j^\m, \qquad\qquad
j^\m(f,g)=(i/\hbar)|g|^{1/2}g^{\m\n}\Bigl(\overline{f}\p_\n g -
(\p_\n \overline{f})g\Bigr).\label{KGip}\eeq
This bracket is sometimes called the {\it Klein-Gordon inner product}, and $\la f,f\ra$ the {\it Klein-Gordon norm} of $f$.
The current density $j^\m(f,g)$ is divergenceless ($\p_\m j^\m=0$)
when the functions $f$ and $g$ satisfy the KG equation
(\ref{scalareom}), hence
the value of the integral in (\ref{KGip}) is independent of the
spacelike surface $\S$ over which it is evaluated, provided the
functions vanish at spatial infinity. The KG inner
product satisfies the relations
\beq \overline{\la f,g\ra}=-\la \overline{f},\overline{g}\ra = \la g,f\ra,
\qquad\qquad \la f,\overline{f}\ra=0 \label{bracketrelns}\eeq
Note that it is not positive definite.

\subsection{Hilbert space}
\label{Hilbert}
At this point it is common to expand the field operator in
modes and to associate annihilation and creation operators with
modes, in close analogy with the harmonic oscillator
(\ref{xoperator}), however instead I will begin with individual
wave packet solutions. My reason is that in some situations
there is no particularly natural set of modes, and none is
needed to make physical predictions from the theory. (An
illustration of this statement will be given in our treatment of 
the Hawking effect.) A mode
decomposition is a basis in the space of solutions, and has no
fundamental status.

In analogy with the harmonic oscillator case (\ref{pluckout}),
we define the annihilation operator associated with a complex
classical solution $f$ by the bracket of $f$ with the field
operator $\fphi$:
\beq a(f)=\la f,\fphi\ra \label{a(f)}\eeq
Since both $f$ and $\fphi$ satisfy the wave equation, $a(f)$ is
well-defined, independent of the surface on which the bracket
integral is evaluated. It follows from the above definition and
the hermiticity of $\fphi$ that the hermitian conjugate of
$a(f)$ is given by
\beq \ad(f)=-a(\overline{f}).\label{ad(f)}\eeq
The canonical commutation relation (\ref{ccr}) together with
the definition of the momentum (\ref{momentum}) imply that
\beq [a(f),\ad(g)]=\la f,g\ra.\label{aadcomm}\eeq
The converse is also true, in the sense that if (\ref{aadcomm})
holds for all solutions $f$ and $g$, then the canonical
commutation relation holds. Using (\ref{ad(f)}), we immediately
obtain the similar relations
\beq [a(f),a(g)]=-\la f,\overline{g}\ra,\qquad\qquad
[\ad(f),\ad(g)]=-\la \overline{f},g\ra \label{aacomm}\eeq

If $f$ is a positive norm
solution with unit norm $\la f,f\ra=1$, then $a(f)$ and
$\ad(f)$ satisfy the usual commutation relation for the raising
and lowering operators for a harmonic oscillator,
$[a(f),\ad(f)]=1$. Suppose now that $|\Psi\ra$ is a normalized
quantum state satisfying $a(f)|\Psi\ra=0$. Of course this
condition does not specify the state, but rather only one
aspect of the state. Nevertheless, for each $n$, the state
$|n,\Psi\ra=(1/\sqrt{n!})(\ad(f))^n|\Psi\ra$ is a normalized
eigenstate of the number operator $N(f)=\ad(f)a(f)$ with
eigenvalue $n$. The span of all these states defines a Fock
space of $f$-wavepacket ``$n$-particle excitations" above the
state $|\Psi\ra$.

If we want to construct the {\it full} Hilbert space of the
field theory, how can we proceed? We should find a
decomposition of the space of complex solutions to the wave
equation $\cal S$ into a direct sum of a positive norm subspace
${\cal S}_p$ and its complex conjugate $\overline{{\cal S}_p}$,
such that all brackets between solutions from the two subspaces
vanish. That is, we must find a direct sum decomposition
\beq{\cal S}={\cal S}_p \oplus\overline{{\cal S}_p} 
\label{decomp}
\eeq
such that
\begin{eqnarray}
 \la f,f\ra &>& 0 \qquad \forall f\in {\cal S}_p\\
\la f,\overline{g}\ra &=& 0 \qquad \forall f,g\in {\cal S}_p.
\label{decompfg}
\end{eqnarray}
The first condition implies that each $f$ in ${\cal S}_p$ can
be scaled to define its own harmonic oscillator sub-albegra as
in the previous paragraph. The second condition implies,
according to (\ref{aacomm}), that the annihilators and creators
for $f$ and $g$ in the subspace ${\cal S}_p$ commute amongst
themselves: $[a(f),a(g)]=0=[\ad(f),\ad(g)]$. 

Given such a decompostion a total Hilbert space for the field theory can be defined 
as the space of finite
norm sums of possibly infinitely many states of the form
$\ad(f_1)\cdots\ad(f_n)|0\ra$, where 
$|0\ra$ is a state such that $a(f)|0\ra=0$ for {\it all} $f$ in
${\cal S}_p$, and all $f_1,\dots,f_n$ are in
${\cal S}_p$. The state $|0\ra$ is called a
{\it Fock vacuum}. It depends on the
decomposition (\ref{decomp}), and in general is not the ground
state (which is not even well defined unless the background metric
is globally static). The representation
of the field operator on this Fock space is hermitian and
satisfies the canonical commutation relations.  

\subsection{Flat spacetime}
\label{flat}

Now let's apply the above generalities to the case of a massive
scalar field in flat spacetime. In this setting
a natural decomposition of the space of solutions is
defined by positive and negative frequency 
with respect to a Minkowski time translation, and the corresponding
Fock vacuum is the ground state.
I summarize briefly since this is standard flat spacetime quantum field theory. 

Because of the infinite volume of space, plane wave solutions are of
course not normalizable. To keep the physics straight and the language
simple it is helpful
to introduce periodic boundary conditions, so that space
becomes a large three-dimensional torus with circumferences $L$
and volume $V=L^3$. The allowed wave vectors are then 
${\bf k}= (2\pi/L){\bf n}$, where the components of the vector
$\bf n$ are integers. In the end we can always take the limit
$L\rightarrow \infty$ to obtain results for local quantities that 
are insensitive to this formal compactification. 

A complete set
of solutions (``modes") to the classical wave equation (\ref{scalareom})
with $\Box$ the flat space d'Alembertian and $R=0$ is given by
\beq f_{\bf k}(t,{\bf x})=\sqrt{\frac{\hbar}{2V\o({\bf k})} }
e^{-i\o({\bf k})t}e^{i{\bf k}\cdot{\bf x}} \label{fk}\eeq
where
\beq \o({\bf k})=\sqrt{{\bf k}^2 + m^2},\eeq
together with the solutions obtained by replacing the positive
frequency $\o({\bf k})$ by its negative, $-\o({\bf k})$. The
brackets between these solutions satisfy
\bea \la f_{\bf k}, f_{\bf l}\ra &=& \d_{{\bf k}, {\bf l}}\label{ff}\\
\la \overline{f}_{\bf k}, \overline{f}_{\bf l}\ra &=& -\d_{{\bf k}, {\bf l}}\label{fbfb}\\
\la f_{\bf k}, \overline{f_{\bf l}}\ra &=& 0, \label{ffb}\eea
so they provide an orthogonal decomposition of the solution
space into positive norm solutions and their conjugates
as in (\ref{decompfg}), with ${\cal S}_p$ the space spanned
by the positive frequency modes $f_{\bf k}$.
As described
in the previous subsection this provides 
a Fock space representation. 

If we define the
annihilation operator associated to $f_{\bf k}$ by
\beq a_{\bf k}=\la f_{\bf k},\fphi\ra,\eeq
then the field operator has the expansion
\beq \fphi=\sum_\bk\; \left(f_{\bf k}\, a_{\bf k} +
\overline{f}_{\bf k}\, \ad_{\bf k}\right).
\label{phiexpansion}\eeq
Since the individual solutions $f_{\bf k}$ have positive frequency,
and the Hamiltonian is a sum over the  contributions
from each $\bf k$ value,
our previous  discussion of the single oscillator shows that
the vacuum state defined by
\beq a_{\bf k}|0\ra=0 \qquad  \eeq
for all ${\bf k}$ is in fact the ground state of the Hamiltonian. 
The states
\beq \ad_{\bf k}|0\ra \eeq
have momentum $\hbar{\bf k}$ and energy $\hbar\o({\bf k})$, and
are interpreted as single particle states. States of the form
$\ad_{\bf k_1}\cdots\ad_{\bf k_n}|0\ra$ are interpreted as $n$-particle states.

Note that although the field Fourier component 
$\fphi_{\bf k} = f_{\bf k}\, a_{\bf k} +
\overline{f}_{-\bf k}\, \ad_{-\bf k}$ has zero 
mean in the vacuum state, like the harmonic 
oscillator position it undergoes 
``zero-point fluctuations" characterized by 
\beq
\la0|\fphi_{\bf k}^\dagger\fphi_{\bf k}|0\ra
=|f_{-\bf k}|^2=\frac{\hbar}{2V\o({\bf k})},
\label{fieldzp}
\eeq
which is entirely analogous to the oscillator result (\ref{zpf}).

\subsection{Curved spacetime, ``particles", and stress tensor}

In a general curved spacetime setting there is no analog of
the preferred Minkowski vacuum and definition of particle
states. However, it is clear that we can import these notions
locally in an approximate sense if the wavevector and frequency
are high enough compared to the inverse radius of curvature.
Slightly more precisely, we can expand the metric in Riemann
normal coordinates about any point $x_0$:
\beq g_{\m\n}(x)
=\eta_{\m\n}+\frac{1}{3}R_{\m\n\a\b}(x_0)(x-x_0)^\a(x-x_0)^\b +
O((x-x_0)^3).\eeq
If ${\bf k}^2$ and $\o^2({\bf k})$ are much larger than any
component of the Riemann tensor $R_{\m\n\a\b}(x_0)$ in this
coordinate system then it is clear that the flat space
interpretation of the corresponding part of 
Fock space will hold to a good
approximation, and in particular a particle detector will
respond to the Fock states as it would in flat spacetime. This
notion is useful at high enough wave vectors locally in any
spacetime, and for essentially all wave vectors asymptotically
in spacetimes that are asymptotically flat in the past or the
future or both.

More generally, however, the notion of a ``particle" is
ambiguous in curved spacetime, and one should use field
observables to characterize states. One such observable 
determines how a ``particle detector" coupled to the field would
respond were it following some particular worldline in
spacetime and the coupling were adiabatically turned on and off
at prescribed times. For example the transition probability 
of a point monopole 
detector is determined in lowest order perturbation theory by
the two-point function $\la\Psi|\fphi(x)\fphi(x')|\Psi\ra$ evaluated
along the worldline\cite{Unruh:ga,DeWittEC} 
of the detector. 
(For a careful discussion of the regularization required
in the case of a point detector see \cite{Schlicht:2003iy}.)
Alternatively this quantity---along with the higher order correlation
functions---is itself a probe of the state of the field. 

Another example of a field observable is the 
expectation value of the stress energy
tensor, which is the source term in the semi-classical Einstein
equation
\beq
G_{\m\n}=8\pi G\la\Psi| T_{\m\n}(x)|\Psi\ra.
\label{scee}
\eeq
This quantity is infinite
because it contains the product of field operators
at the same point.
Physically, the infinity is due to
the fluctuations of the infinitely many ultraviolet field modes.
For example, the leading order divergence of the energy
density can be 
attributed to the zero-point energy of the field fluctuations,
but there are subleading divergences as well.
We have no time here to properly go into this subject,
but rather settle for a few brief comments.

One way to make sense of the expectation value is 
via the {\it difference} between its values in two different states.
This difference is well-defined (with a suitable
regulator) and finite for any two states sharing the 
same singular short distance behavior. The result depends of course
upon the comparison state.
Since the divergence is associated with the 
very short wavelength modes, it might seem that to
uniquely define the expectation value at a point $x$ it should
suffice to just subtract the infinities for a state
defined as the vacuum in the local flat spacetime 
approximation at $x$.
This subtraction is ambiguous however, 
even after making it as local as
possible and ensuring local conservation of energy
($\nabla^\m \la T_{\m\n}\ra=0$). 
It defines the expectation value only up to a 
tensor $H_{\m\n}$ constructed locally from the 
background metric with four or fewer derivatives
and satisfying the identity $\nabla^\m H_{\m\n}=0$.
The general such tensor is the variation with respect
to the metric of the invariant functional
\beq \int d^Dx\, \sqrt{|g|}\, \Bigl(c_0+c_1R + c_2 R^2 +c_3
R^{\m\n}R_{\m\n} + c_4R^{\m\n\r\s}R_{\m\n\r\s}\Bigr).
\label{Haction}
\eeq
(In four spacetime dimensions the last term can be rewritten 
as a combination of the first two and a total divergence.) 
Thus $H_{\m\n}$ is a combination of $g_{\m\n}$, the 
Einstein tensor $G_{\m\n}$, and curvature squared terms.
In effect, the ambiguity $-H_{\m\n}$ is added to the metric side of the semi-classical field equation, where it renormalizes the cosmological constant and Newton's constant, and introduces curvature squared terms.  

A different approach is to define the expectation value
of the stress tensor via the metric variation of the renormalized 
effective action, which posesses ambiguities of the same form as (\ref{Haction}). Hence the two approaches agree.

\subsection{Remarks}

\subsubsection{Continuum normalization of modes}
Instead of the ``box normalization" used above 
we could ``normalize"  the solutions $f_{\bf k}$ (\ref{fk})
with the factor $V^{-1/2}$ replaced by $(2\pi)^{-3/2}$. 
Then the Kronecker $\d$'s  in
(\ref{ff}, \ref{fbfb}) 
would be replaced by Dirac $\d$-functions and 
the discrete sum over momenta in (\ref{phiexpansion}) 
would be an integral over $\bk$. In this case the
annihilation and creation operators would satisfy 
$[a_\bk,\ad_{\bl}]=\d^3(\bk,\bl)$.

\subsubsection{Massless minimally coupled zero mode}
The massless minimally coupled
case $m=0$, $\xi=0$ has a peculiar feature. The
spatially constant function $f(t,{\bf x})=c_0 + c_1t$ is a
solution to the wave equation that is not included among the
positive frequency solutions $f_{\bf k}$ or their conjugates. 
This ``zero mode" must be quantized as well, but
it behaves like a free particle rather than like a harmonic
oscillator. In particular, the state of lowest energy would
have vanishing conjugate field momentum and hence would be
described by a Schr\"odinger wave function $\psi(\fphi_0)$
that is totally
delocalized in the field amplitude $\fphi_0$. Such a wave function would 
be non-normalizable as a quantum state, just as
any momentum eigenstate of a non-relativistic particle
is non-normalizable. Any normalized state would be described by a
Schr\"odinger wavepacket that would spread in $\fphi_0$
like a free particle spreads in position space, and would
have an expectation value $\la\fphi_0\ra$ growing linearly in time. 
This suggests 
that no time-independent state exists. That is indeed true in the
case where the spatial directions are compactified on a torus
for example, so that the modes are discrete and the zero mode
carries as much weight as any other mode. In non-compact
space one must look more closely. It turns out
that in
$1+1$ dimensions the zero mode continues to preclude a time
independent state (see e.g. \cite{Ford:1985qh}, although the
connection to the behavior of the zero mode is not made there), 
however in higher dimensions it does not,
presumably because of the extra factors of $k$ in the measure
$k^{D-2}dk$. A version of the same issue arises in deSitter
space, where no deSitter invariant state exists for the
massless, minimally coupled field~\cite{Allen:ux}. That is true in
higher dimensions as well, however, which may be related to the
fact that the spatial sections of deSitter space are compact.

\sect{Particle creation}
We turn now to the subject of particle creation in 
curved spacetime (which would more appropriately be 
called ``field excitation", but we use the standard term). 
The main applications
are to the case of expanding cosmological spacetimes and to 
the Hawking effect for black holes. To begin with I will discuss the
analogous effect for a single harmonic oscillator, which already contains 
the essential elements of the more complicated cases. The results will
then be carried over to the cosmological setting. The following 
section then takes up the subject of the Hawking effect.

\subsection{Parametric excitation of a harmonic oscillator}
A quantum field in a time-dependent background spacetime can
be modeled in a simple way by a harmonic oscillator whose frequency $\o(t)$ is a
given function of time. The equation of motion is then (\ref{ho(t)}),
\beq
\ddot{x}+\o^2(t)\, x = 0.
\label{hoeom}
\eeq
We consider the situation where the frequency is asymptotically constant,
approaching $\o_{in}$ in the past and $\o_{out}$ in the future.
The question to be answered is this: if the oscillator
starts out in the ground state $|0_{in}\ra$ 
appropriate to $\o_{in}$ as $t\rightarrow -\infty$,
what is the state as  $t\rightarrow +\infty$? More precisely, in the Heisenberg
picture the state does not evolve, so what we 
are really asking is how is the state $|0_{in}\ra$ expressed as
a Fock state in the out-Hilbert space  appropriate to $\o_{out}$?
It is evidently not the same as the ground state $|0_{out}\ra$ of the
Hamiltonian in the asymptotic future.

To answer this question we need only relate the annihilation and creation
operators associated with the {\it in} and {\it out} normalized 
positive frequency modes $f_{\stackrel{in}{out}}(t)$,
which are solutions to (\ref{hoeom}) with the asymptotic behavior
\beq
f_{\stackrel{in}{out}}(t)\stackrel{t\rightarrow\mp\infty}{\longrightarrow}
\sqrt{\frac{\hbar}{2m\o_{\stackrel{in}{out}}}}\, \exp(-i\o_{\stackrel{in}{out}} t).
\eeq
Since the equation of motion is second order in time derivatives 
it admits a two-parameter family of
solutions, hence there must exist complex constants $\a$ and $\b$ such that
\beq
f_{out}=\a f_{in} + \b \bar{f_{in}}.
\label{Bogmode}
\eeq
The normalization condition $\la f_{out},f_{out}\ra=1$ implies that 
\beq
|\a|^2 - |\b|^2 = 1.
\label{Bognorm}
\eeq
The {\it out} annihilation operator is given (see (\ref{pluckout})) by 
\bea
a_{out} &=& \la f_{out},x\ra\\ 
&=& \la \a f_{in} + \b \bar{f_{in}},x\ra\\ 
&=& \a a_{in}-\bar{\b}a_{in}^\dagger.
\label{aout}
\eea
This sort of linear relation between two sets of annihilation and
creation operators (or the corresponding solutions)
is called a {\it  Bogoliubov transformation}, and the 
coefficients are the {\it Bogoliubov coefficients}. 
The mean value of the {\it out} number operator $N_{out}=a_{out}^\dagger a_{out}$
is nonzero in the  state $|0_{in}\ra$: 
\beq
\la0_{in}|N_{out}|0_{in}\ra=|\b|^2.
\label{Nout}
\eeq
In this sense the time dependence of $\o(t)$ excites the
oscillator, and $|\beta|^2$ characterizes the excitation number. 

To get a feel for the Bogoliubov coefficient $\b$ let us 
consider two extreme cases,  adiabatic and sudden.

\subsubsection{Adiabatic transitions and ground state}
The adiabatic case corresponds to a situation in which the
frequency is changing very slowly compared to the period 
of oscillation,
\beq
\frac{\dot{\o}}{\o}\ll\o.
\eeq
In this case there is almost no excitation, 
so $|\b|\ll1$. 
Generically in the adiabatic case the Bogoliubov 
coefficient is exponentially 
small, $\b\sim \exp(-\o_0 T)$, where $\o_0$ is a typical frequency 
and $T$ characterizes the time scale for the variations in the frequency.
 
In the Schr\"odinger picture, one can say that during  
an adiabatic change of $\o(t)$ the state continually
adjusts to remain close to the instantaneous {\it adiabatic ground state}.
This state at time $t_0$ is the one annihilated by the lowering operator
$a(f_{t_0})$ defined by the solution $f_{t_0}(t)$ satisfying the initial conditions
at $t_0$ corresponding to the ``instantaneous positive frequency solution",
\bea
{f}_{t_0}(t_0)&=&\sqrt{\hbar/2m\o(t_0)}\\
\dot{f}_{t_0}(t_0)&=&-i\o(t_0)f_{t_0}(t_0).
\label{adiabatground}
\eea
The instantaneous adiabatic ground state is sometimes
called the ``lowest order adiabatic ground state at $t_0$". One
can also consider higher order adiabatic ground states
as follows (see e.g. \cite{BD,Fulling89}).
A function of the WKB form 
$(\hbar/2mW(t))^{1/2}\exp(-i\int^t W(t')\, dt')$ is 
a normalized solution to (\ref{hoeom}) provided 
$W(t)$ satisfies a certain second order differential
equation. That equation can be solved iteratively, yielding 
an expansion $W(t)=\o(t) +\cdots$, where the subsequent 
terms involve time derivatives of $\o(t)$. The lowest order
adiabatic ground state at $t_0$ is defined using the solution 
whose initial conditions (\ref{adiabatground}) 
match the lowest order truncation of the expansion for $W(t)$. A higher order
adiabatic ground state is similarly defined using a higher order truncation.

\subsubsection{Sudden transitions}
The opposite extreme is the sudden one, in which $\o$ changes instantaneously
from $\o_{in}$ to $\o_{out}$ at some time $t_0$.  We can then find the Bogoliubov 
coefficients using (\ref{Bogmode}) and its first derivative at $t_0$. 
For $t_0=0$ the result is 
\bea
\a&=&\frac{1}{2}\left(\sqrt{ \frac {\o_{in}} {\o_{out}} } +
\sqrt{\frac{\o_{out}}{\o_{in}} }\right)\\
\b&=&\frac{1}{2}\left(\sqrt{\frac{\o_{in}}{\o_{out}}}-
\sqrt{\frac{\o_{out}}{\o_{in}}}\right).
\eea
(For $t_0\ne0$ there are extra phase factors in these solutions.)
Interestingly, the amount of excitation is precisely the same if
the roles of $\o_{in}$ and $\o_{out}$ are interchanged. 
For an example, consider the case where $\o_{out}=4\o_{in}$,
for which $\a=5/4$ and  $\b=-3/4$. In this case the expectation
value (\ref{Nout}) of the out number operator is 9/16, so there
is about ``half an excitation".

\subsubsection{Relation between {\it in} and {\it out} ground states \& the 
squeeze operator}
The expectation value of $N_{out}$ is only
one number characterizing the relation between
$|O_{in}\ra$ and the out states. We shall now
determine the complete relation 
\beq
|O_{in}\ra=\sum_n c_n |n\ra_{out}
\eeq
where the states $|n\ra_{out}$ are eigenstates
of $N_{out}$ and the $c_n$ are constants.

One can find $a_{in}$ in terms of  $a_{out}$ and $\ad_{out}$ 
by combining (\ref{Bogmode}) with its complex conjugate to solve for
$f_{in}$ in terms of $f_{out}$ and $\bar{f}_{out}$. In 
analogy with (\ref{aout}) one then finds 
\beq
a_{in}  = \a a_{out}+\bar{\b}a_{out}^\dagger.
\label{Bogin}
\eeq
Thus the defining condition 
$a_{in}|0_{in}\ra=0$ implies
\beq
a_{out}|0_{in}\ra=-\frac{\bar{\b}}{\a}\, \ad_{out}|0_{in}\ra.
\label{gnd}
\eeq
A transparent way to solve this is to 
note that the commutation relation 
$[a_{out},\ad_{out}]=1$ suggests the formal analogy
$a_{out} = \partial/\partial \ad_{out}$. This casts
(\ref{gnd}) as a first order ordinary differential
equation, with solution
\bea
|0_{in}\ra&=&{\cal N}\exp\left[-\left(\frac{\bar{\b}}{2\a}\right)
\ad_{out}\ad_{out}\right]|0_{out}\ra\label{squeezedgnd}\\
&=& {\cal N}\sum_n \frac{\sqrt{2n!}}{n!}\left(-\frac{\bar{\b}}{2\a}\right)^n\, |2n\ra_{out}.
\label{ininout}
\eea
Note that the state $|0_{out}\ra$ on which the exponential operator acts is annihilated
by $a_{out}$, so there is no extra term from $\ad_{out}$-dependence.

The state $|0_{in}\ra$ thus contains only even numbered excitations when 
expressed in terms of the out number eigenstates. 
The normalization constant ${\cal N}$ is given by 
\beq
|{\cal N}|^{-2}=\sum_n \frac{2n!}{(n!)^2}\left|\frac{\bar{\b}}{2\a}\right|^{2n}.
\eeq
For large $n$ the summand approaches $|\bar{\b}/\a|^{2n}=\left[|\b|^2/(|\b|^2+1)\right]^n$, 
where the relation (\ref{Bognorm}) was used in the last step.
The sum therefore converges, and can be 
evaluated\footnote{Alex Maloney pointed out that it
can be evaluated by expressing the binomial coefficient $2n!/(n!)^2$
as the contour integral $\oint(dz/2\pi i z)(z+1/z)^{2n}$ and interchanging
the order of the sum with the integral.} to yield 
\beq
|{\cal N}|=\Bigl(1-|\b/\a|^2\Bigr)^{1/4} = |\a|^{-1/2}.
\label{N1mode}
\eeq

An alternate way of describing $|0_{in}\ra$ in the out Hilbert space
is via the {\it squeeze operator}
\beq
S = \exp\Bigl[\frac{z}{2}\ad\ad-\frac{\bar{z}}{2}aa\Bigr].
\label{squeezeop}
\eeq
Since its exponent is anti-hermitian, $S$ is unitary.
Conjugating $a$ by $S$ yields 
\beq
S^\dagger a S = \cosh |z| \, a +\sinh |z| \frac{z}{|z|} \, a^\dagger.
\eeq
This has the form of the Bogoliubov transformation
(\ref{Bogin}) with $\a=\cosh|z|$ and $\b=\sinh |z|({z}/{|z|})$.
With $a_{out}$ in place of $a$ in $S$ this gives
\beq
a_{in} = S^\dagger a_{out} S.
\eeq
The condition $a_{in}|0_{in}\ra=0$ thus implies
$a_{out} S|0_{in}\ra=0$, so evidently
\beq
|0_{in}\ra=S^\dagger |0_{out}\ra
\label{squeezedstate}
\eeq
up to a constant phase factor.
That is, the {\it in} and {\it out} ground states 
are related by the action of the
squeeze operator $S$. Since $S$ is 
unitary, the right hand side of (\ref{squeezedstate})
is manifestly normalized.

\subsection{Cosmological particle creation}

We now apply the ideas just developed to a free 
scalar quantum field satisfying the KG equation
(\ref{scalareom}) 
in a homogeneous isotropic spacetime. The case when the spatial 
sections are flat is slightly simpler, and it is already
quite applicable, hence we restrict to that case
here.  

The spatially flat Robertson-Walker (RW) line element takes the form:
\beq ds^2=dt^2 - a^2(t) dx^i dx^i = a^2(\eta)(d\eta^2 -dx^i
dx^i)
\label{RW}
\eeq
It is conformally flat, as are all RW metrics.
The coordinate $\eta=\int dt/a(t) $ is called the
{\it conformal time}, to distinguish it from the 
proper time $t$ of the isotropic observers. 
The d'Alembertian $\Box$ for this metric 
is given by
\beq 
\Box = \p_t^2 + (3\dot{a}/a)\p_t - a^{-2}\p_{x^{i}}^2,
\label{cosmodal}
\eeq
where the dot stands for $\partial/\partial t$.
The spatial translation symmetry allows the spatial
dependence to be separated from the time dependence. 
A field
\beq 
u_k(\bx,t) = \z_\bk(t)e^{i\kdx} 
\label{zetamode}
\eeq 
satisfies the
field equation (\ref{scalareom}) 
provided $\z_\bk(t)$ satisfies an equation similar to 
that of a damped harmonic
oscillator but with time-dependent damping coefficient 
$3\dot{a}/a$ and time-dependent 
frequency $a^{-2}k^2+m^2+\xi R$. It is worth emphasizing
that in spite of the ``damping", the field equation is Hamiltonian,
and the Klein-Gordon norm (\ref{KGip}) of any solution is conserved in the evolution.  

The field equation can be put into the form of an undamped oscillator
with time-dependent frequency by using the conformal time $\eta$ 
instead of $t$ and factoring out an appropriate  power of the
conformal factor $a^2(\eta)$:
 \beq 
\z_\bk = a^{-1}\, \chi_\bk.
\label{chi}
\eeq 
The function $u_{\bk}$ satisfies the field equation if and only if
\beq
\chi_\bk'' + \o^2(\eta)\chi_\bk=0,
\label{chieom}
\eeq
where the prime stands for $d/d\eta$ and 
\beq \o^2(\eta) = k^2+ m^2a^2 - (1-6\xi)(a''/a). \eeq
(In the special case of conformal coupling
$m=0$ and $\xi=1/6$, this becomes the time-independent harmonic
oscillator, so that case is just like flat spacetime. All effects of the
curvature are then incorporated by the prefactor $a(\eta)^{-1}$ in (\ref{chi}).)

The  $u_\bk$ are orthogonal in the Klein-Gordon  inner product (\ref{KGip}),
and they are normalized\footnote{The two inverse factors of $a$
coming from (\ref{chi}) are cancelled by the $a^3$ in the volume element
and the $a^{-1}$ in the relation $\partial/\partial t = a^{-1}\partial/\partial \eta$} 
provided the $\chi_\bk$ have unit norm:
\beq
\la u_\bk,u_\bl\ra=\d_{\bk,\bl}\iff (iV/\hbar)(\bar{\chi}_\bk\chi_\bk'-\bar{\chi}_\bk'\chi_\bk)=1.
\eeq
Note that the relevant norm for the $\chi_\bk$ differs from that for the harmonic oscillator
(\ref{bracket}) only by the replacement $m\rightarrow V$, where
$V$ is the $x^i$-coordinate volume of the constant $t$ surfaces.
We also have $\la u_\bk,\bar{u}_\bl\ra=0$
for all $\bk, \bl$, hence these modes provide an orthogonal 
postive/negative norm decomposition  of the space of complex solutions.
As discussed in section \ref{Hilbert}, this yields a corresponding Fock space
representation for the field operators.
The field operator can be expanded in terms of the corresponding annihilation
and creation operators:
\beq
\fphi (\bx,t)=\sum_{\bk} \; \Bigl(u_\bk(\bx,t) \, a_\bk + \bar{u}_\bk(\bx,t) \, \ad_\bk\Bigr) .
\eeq

Consider now the special case where there is no time dependence in the
past and future, $a(\eta)\rightarrow$ constant. The {\it in} and {\it out} ``vacua" 
are the ground states of the Hamiltonian at early and late times, and are
the states annihilated by the $a_\bk$ associated with the $u_\bk^{in,out}$
constructed with early and late time
positive frequency modes $\chi_\bk^{in,out}$, as explained in section \ref{flat}:
\beq
\chi_\bk^{\stackrel{in}{out}}(\eta)\stackrel{\eta\rightarrow\mp\infty}{\longrightarrow}
\sqrt{\frac{\hbar}{2V\o_{\stackrel{in}{out}}}}\, \exp(-i\o_{\stackrel{in}{out}} \eta).
\eeq
The Bogoliubov transformation now takes the form
\beq 
u_\bk^{\rm out} =\sum_{\bk'}\; \Bigl(\a_{\bk\bk'}u_{\bk'}^{in} + 
\b_{\bk\bk'}\bar{u}_{\bk'}^{in}\Bigr).
\label{Bogcosmo}
\eeq
Matching the coefficients of $\exp(i\bk\cdot\bx)$, we see that 
\beq
\a_{\bk\bk'}=\a_{k}\d_{\bk,\bk'}, \qquad\qquad \b_{\bk\bk'}=\b_{k}\d_{\bk,-\bk'},
\eeq
i.e. the Bogoliubov coefficients mix only modes of wave vectors $\bk$ and $-\bk$,
and they depend only upon the magnitude of the wavevector on account of rotational symmetry (eqn. (\ref{chieom}) for $\chi_\bk$ does not depend on the direction of $\bf k$).
The normalization condition on $u_\bk^{\rm out}$ implies
\beq
|\a_k|^2 - |\b_k|^2 = 1.
\label{Bognormk}
\eeq

As in the harmonic oscillator example (\ref{Nout}), if the state is the {\it in}-vacuum, 
then the expected excitation level of the $\bk$ out-mode,
i.e. the average number of particles in that mode, is given by 
\beq
\la 0_{in}|N_{\bk}^{out}|0_{in}\ra=|\b_k|^2.
\label{Nkaverage}
\eeq
To convert this statement into one about 
particle density, we sum over $\bk$ and divide by the physical spatial volume
$V_{phys}=a^3 V$,
which yields the number density of particles. Alternatively one can work with 
the continuum normalized modes. The relation between the discrete and continuous 
sums is 
\beq
\frac{1}{V_{phys}}\sum_\bk \longleftrightarrow \frac{1}{(2\pi a)^3}\int d^3\bk.
\eeq
the number density of out-particles is thus
\beq
n^{out}=  \frac{1}{(2\pi a)^3}\int d^3\bk |\b_k|^2 .
\label{nout}
\eeq

The mean particle number characterizes only certain aspects of the state.
As in the oscillator example, a full description of the in-vacuum in the out-Fock space
is obtained from the Bogoliubov relation between the corresponding annihilation
and creation operators. From (\ref{Bogcosmo}) we can solve for $u^{in}$ and 
thence find 
\beq
a_\bk^{in} = \a_k\,  a_\bk^{out}+\bar{\b}_k \, \a_{-\bk}^\dagger{}^{out},
\eeq
whence
\beq 
a_\bk^{out} |0_{in}\ra = -\frac{\bar{\b}_k}{\a_k} \, a_{-\bk}^\dagger{}^{out}  |0_{in}\ra,
\eeq
which can be solved to find
\beq
|0_{in}\ra=\left(\prod_{\bk'} {\cal N_\bk'}\right)\exp\left[-\sum_{\bk}\left(\frac{\bar{\b_k}}{2\a_k}\right)
a_{\bk}^\dagger{}^{out}a_{-\bk}^\dagger{}^{out}\right]|0_{out}\ra
\label{squeezedvac}
\eeq
where the $\cal N_\bk'$ are normalization factors.
This solution 
is similar to the corresponding expression (\ref{ininout}) for the harmonic oscillator and it can be found by a similar method. Although the operators 
$a_{\bk}^\dagger{}^{out}$ and $a_{-\bk}^\dagger{}^{out}$
are distinct, each product appears twice in the sum, once for $\bf k$ and once for $-{\bf k}$, hence the factor of 2 in the denominator of the exponent is required. 
Using the two-mode analog of the squeeze operator (\ref{squeezeop})
the state (\ref{squeezedvac}) can also be written in a manifestly 
normalized fashion analogous to (\ref{squeezedstate}). It is sometimes
called a {\it squeezed vacuum}.

\subsection{Remarks}

\subsubsection{Momentum correlations in the squeezed vacuum}
The state (\ref{squeezedvac}) can be expressed as a sum of terms
each of which has equal numbers of $\bk$ and $-\bk$ excitations.
These degrees of freedom are thus entangled in the state, in such
a way as to ensure zero total momentum. 
This is required by translation invariance of the states 
$|0_{in}\ra$ and $|0_{out}\ra$,
since momentum is the generator of space translations.

\subsubsection{Normalization of the squeezed vacuum}
The norm sum for the part of the state  (\ref{squeezedvac})
involving $\bk$ and $-\bk$ 
is a standard geometric series, which evaluates 
to $|\a_k|^2$ using (\ref{Bognormk}). 
Hence to normalize the state one can set
${\cal N}_\bk=|\a_k|^{-1/2}$ for all $\bk$, including 
$\bk=0$ as in (\ref{N1mode}). The overall normalization factor 
is a product of infinitely many numbers less than unity. Unless
those numbers converge rapidly enough to unity the state is 
not normalizable. The condition for normalizability is easily seen to
be $\sum_{k}|\b_k|^2<\infty$, i.e. according to (\ref{Nkaverage})
the average total number of excitations must be finite. If it is
not, the state $|0_{in}\ra$ does not lie in the Fock space built on the 
the state  $|0_{out}\ra$. Note that,
although formally unitary, the squeeze operator does not 
act unitarily on the {\it out} Fock space if the corresponding state is 
not in fact normalizable.

\subsubsection{Energy density}
If the scale factor $a$ changes by over a time interval $\D\t$,
then for a massless field dimensional analysis indicates that the 
in vacuum has a resulting energy density $\rho\sim \hbar(\D\t)^{-4}$ after the
change. To see how the formalism produces this, according to (\ref{nout})
we have $\r=(2\pi a)^{-3}\int d^3k |\b_k|^2 (\hbar\o/a)$. The Bogoliubov
coefficent $\b_k$ is of order unity around $k_c/a\sim 1/\D \t$ and decays
exponentially above that. The integral is dominated by the upper limit 
and hence yields the above mentioned result.

\subsubsection{Adiabatic vacuum}
Modes with frequency 
much larger than $a'/a$ see the change of the scale factor
as adiabatic, hence they remain relatively unexcited.
The state that corresponds to the instantaneously defined
ground state, in analogy with (\ref{adiabatground}) for the
single harmonic oscillator, is called the {\it adiabatic vacuum}
at a given time. 

\subsection{de Sitter space}

The special case of de Sitter space is of interest
for various reasons. The first is just its high degree of
symmetry, which makes it a convenient arena for the study
of qft in curved space.  It is the maximally symmetric
Lorentzian space with  (constant) positive curvature.
Maximal symmetry  refers to the number of Killing fields,
which is  the same as for flat spacetime. The Euclidean
version of de Sitter space is just the sphere. 

de Sitter (dS) space has hypersurface-orthogonal 
timelike Killing fields, hence is locally
static, which further simplifies matters, but not to the
point of triviality. The reason is that all such Killing
fields have Killing horizons, null surfaces to which they
are tangent, and beyond which they are spacelike. Hence dS
space serves as a highly symmetric analog of a black hole
spacetime. In particular, a symmetric variant of the
Hawking effect takes place in de Sitter space, as first
noticed by Gibbons and Hawking\cite{GH}. See \cite{AndydS}
for a recent review.

Inflationary cosmology provides another important use of
deSitter space, since during the period of exponential
expansion the spacetime metric is well described by dS
space. In this application the dS line element is usually
written using spatially flat RW coordinates:
\beq ds^2 = dt^2 - e^{2Ht} dx^i dx^i. \eeq
These coordinates cover only half of the global dS space, and they do not make
the existence of a 
time translation symmetry manifest. This takes the conformal form (\ref{RW}) with $\eta= -H^{-1}\exp(-Ht)$ and $a(\eta) = -1/H\eta$. The range of $t$ is $(-\infty,\infty)$ while that of $\eta$ is 
$(-\infty,0)$.

The flat patch of de Sitter space is 
asymptotically static with respect to 
conformal time $\eta$
in the past, since 
$a'/a=-1/\eta\rightarrow0$ as
$\eta\rightarrow -\infty$.
Therefore in the asymptotic past the adiabatic vacuum 
(with respect to positive $\eta$-frequency) defines a
natural initial state.
This is the initial state used in cosmology. In fact it
happens to define a deSitter invariant state, also
known as the Euclidean vacuum or the Bunch-Davies vacuum.

\subsubsection{Primordial perturbations from zero point fluctuations}
\label{primalpert}

Observations of the Cosmic Microwave Background
radiation support the notion that the origin of 
primordial perturbations lies in the quantum 
fluctuations of scalar and tensor metric modes
(see \cite{Brandenberger:2003vk} for a recent review
and \cite{Mukhanov:1990me} for a classic reference.)
The scalar modes arise from (and indeed
are entirely determined by, since the metric has no 
independent scalar degree of freedom) coupling to 
matter.\footnote{The Cosmic Microwave Background 
observations supporting this account of primordial 
perturbations thus amount
to quantum gravity observations of a limited kind.} Let's
briefly discuss how this works for a massless
minimally coupled scalar field,
which is just how these perturbations are described.
First I'll describe the scenario in words, then add a few
equations.

Consider a field mode with a frequency
high compared to the expansion rate $\dot{a}/a$
during the early universe. To be specific let us 
assume this rate to be a constant $H$, i.e. de Sitter
inflation. Such a mode was
presumably in its ground state,  as the prior 
expansion would have redshifted away any 
initial excitation. 
As the universe expanded the
frequency redshifted until it became comparable to the 
expansion rate, at which point the oscillations 
ceased and the field amplitude approached a time-independent
value. Just before it stopped oscillating the field
had quantum zero point fluctuations of its amplitude,
which were then preserved during the further expansion.
Since the amplitude was frozen when the mode had the 
fixed proper wavenumber $H$, it is the same for all modes
apart from the proper volume factor in the mode normalization
which varies with the cosmological time of freezeout.  
Finally after inflation ended the expansion rate dropped 
faster than the wavenumber, hence eventually the 
mode could begin oscillating again when its wavelength
became shorter than the Hubble length $H^{-1}$. 
This provided the seeds
for density perturbations that would then grow by gravitational
interactions. On account of the particular wavevector dependence 
of the amplitude of the frozen spatial fluctuations, the spectrum of 
these perturbations turns out to be scale-invariant (when
appropriately defined).

More explicitly, 
the field equation for a massless minimally coupled
field is given by (\ref{chieom}),
with 
\beq
\o^2(\eta)= k^2 - a''/a= k^2 - 2H^2a^2,
\eeq
where the last equality holds in de Sitter space.
In terms of proper frequency $\o_p=\o/a$ and proper
wavenumber  $k_p=k/a$
we have $\o_p^2=k_p^2 - 2H^2$. The first term is the usual
flat space one that produces oscillations, while the second term 
tends to oppose the oscillations. 
For proper wavenumbers much higher than $H$ the
second term is negligible. The field oscillates while
the proper wavenumber redshifts exponentially. Eventually
the two terms cancel, and the mode stops oscillating.
This happens when
\beq
k_p= \sqrt{2} H.
\eeq
As the wavenumber continues to redshift into the region $k_p\ll H$, 
to a good approximation 
the amplitude satisfies the equation $\chi_{\bf k}''-(a''/a)\chi_{\bf k}=0$,
which has a growing and a decaying solution\footnote{The general 
solution is $c_1 a + c_2 a\int^\eta d\eta/a^2$, which is  
$c_3 \eta^{-1} + c_4 \eta^2$ in de Sitter space.}. 
The growing solution is $\chi_{\bf k}\propto a$, 
which implies that the field mode
$u_{\bf k}$ (\ref{zetamode},\ref{chi}) is constant in time.
(This conclusion is also evident
directly from  the fact that the last term of the wave operator (\ref{cosmodal})
vanishes as $a$ grows.)

The squared amplitude of the fluctuations when frozen is, according to
(\ref{fieldzp}), 
\beq
\la0|\fphi_{\bf k}^\dagger\fphi_{\bf k}|0\ra= 
|u_{-\bk}|^2 \sim \frac{\hbar}{k_p V_p}= \frac{\hbar H^2}{V k^3},
\eeq
where the proper values are used for consistent matching to the 
previous flat space result. This gives rise to the scale invariant
spectrum of density perturbations.

\sect{Black hole evaporation}

The vacuum of  a quantum field is unstable to particle emission
in the presence of a black hole event horizon. This instability is
called the {\it Hawking effect}. Unlike the cosmological particle creation
discussed in the previous section, this effect is not the result of time
dependence of the metric exciting the field oscillators. Rather it is
more like pair creation in an external electric field\cite{Brout:rd}.
(For more discussion of the role of time dependence see
section \ref{collapse}.)
A general introduction to the Hawking effect was given in section 
\ref{hawkintro}. The present section is devoted to a derivation and discussion
related topics.

The historical roots of the Hawking effect lie in 
the classical Penrose process for extracting energy from 
a rotating black hole. We first review that process
and indicate how it led to Hawking's discovery.
Then we turn to the qft analysis.

\subsection{Historical sketch}
The Kerr metric for a rotating black
hole is stationary, but the asymptotic time translation Killing vector
$\chi$ becomes spacelike {\it outside} the event horizon. 
The region where it is spacelike is called the {\it ergoregion}. 
The conserved Killing energy for a particle with four-momentum $p$
is $E=\chi\cdot p$. Physical particles have future pointing timelike 
4-momenta, hence $E$ is positive provided $\chi$ is also future timelike. 
Where $\chi$ is spacelike however, some physical 4-momenta have negative Killing energy.

In the Penrose process, a particle of energy $E_0>0$ is sent 
into the ergoregion of a rotating black hole where it breaks up 
into two pieces with Killing energies $E_1$ and $E_2$, so that 
$E_0=E_1+E_2$. If $E_2$ is arranged to be negative, then $E_1>E_0$,
that is, more energy comes out than entered. 
The black hole absorbs the negative energy $E_2$ and thus
loses mass. It also loses angular momentum,
hence the process in effect extracts the rotational energy of 
the black hole. 

The Penrose process is maximally efficient and reversible if
the horizon area is unchanged. That condition is achievable in
the limit that the absorbed particle enters the black hole on a trajectory
tangent to one of the null generators of the horizon. This 
role of the horizon area in governing efficiency of energy extraction
exhibits the analogy between area and entropy. Together with Bekenstein's
information theoretic arguments, it gave birth to the subject of
black hole thermodynamics.

When a field scatters from a rotating black hole
a version of the Penrose process called
superradiant scattering can occur. 
The analogy with stimulated 
emission suggests that quantum fields should exhibit 
spontaneous emission from a rotating black hole. 
To calculate this emission rate from an ``eternal"
spinning black hole one must specify a condition on the
state of the quantum field that determines what emerges
from the past horizon. In order to avoid this unphysical
specification Hawking considered instead
a black hole that forms from collapse, which has no past
horizon. In this case only  initial conditions 
before the collapse need be specified. 

Much to his surprise,
Hawking found that for any initial state, even 
a {\it non-rotating} black hole will spontaneously
emit radiation. This {\it Hawking effect}
is a pair creation process in which one member of the pair
lies in the ergoregion inside the horizon and has negative
energy, while the other member lies outside 
and escapes to infinity with positive energy.
The Hawking radiation emerges in a steady flux
with a thermal spectrum at the temperature
$T_H=\hbar\kappa/2\pi$. The surface gravity $\kappa$
had already been seen to play the role of temperature 
in the classical first law of black hole mechanics, which thus
rather remarkably presaged the quantum Hawking effect.

\subsection{The Hawking effect}

Two different 
notions of ``frequency" are relevant to this
discussion. One is the ``Killing frequency", which
refers to time dependence with respect to 
the time-translation symmetry of the background
black hole spacetime. In the asymptotically flat
region at infinity the Killing frequency agrees with 
the usual frequency defined by the Minkowski observers
at rest with respect to the black hole. The other 
notion is ``free-fall frequency" defined by an observer
falling across the event horizon. Since the Killing flow is
tangent to the horizon, Killing frequency
there is very different from free-fall frequency, and that
distinction lies at the heart of the Hawking effect.

\subsubsection{Average number of outgoing particles}

\begin{figure}[htb]
\vbox{ \vskip 8 pt
\centerline{\includegraphics[width=2.7in]{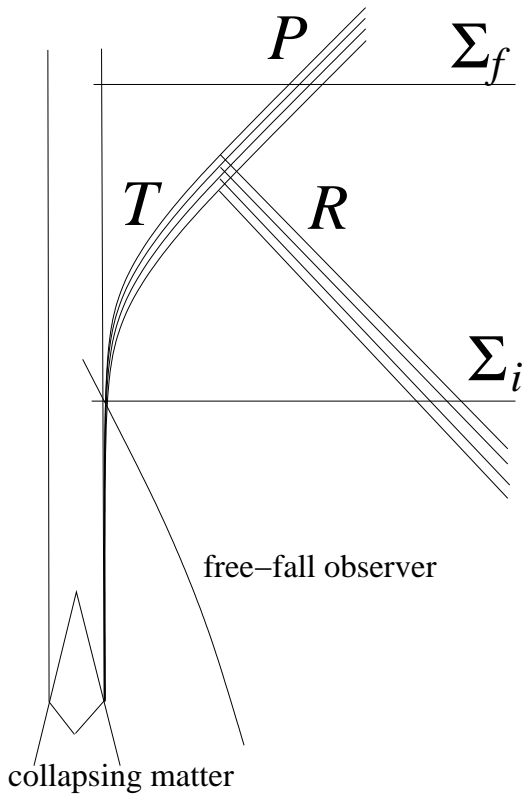}}
\caption{ \label{collapsefig}
\small Spacetime diagram of black hole formed by 
collapsing matter. The outgoing wavepacket $P$ splits
into the transmitted part $T$ and reflected part $R$
when propagated backwards in time. The two surfaces 
$\Sigma_{f,i}$ are employed for evaluating the Klein-Gordon
inner products between the wavepacket and the field
operator. Although $P$, and hence $R$ and $T$, have purely
positive Killing frequency, the free-fall observer 
crossing $T$ just outside the horizon sees both positive 
and negative frequency components with respect to his proper
time.
\smallskip}
}\end{figure}

The  question to be answered is this: if a black
hole forms from collapse
with the quantum field in any `regular' state $|\Psi\ra$,
 then at late times, long after the collapse, what will be the 
average particle number and other observables
for an outgoing positive Killing frequency 
wavepacket $P$ of a  quantum field
far from the black hole?
We address this question here for the case of a noninteracting
scalar field and a static (non-rotating) 
black hole  At the end we make some brief remarks
about generalizations. Figure \ref{collapsefig} depicts
the various ingredients in the following discussion.

To begin with we evaluate the expectation value
$\la\Psi|N(P)|\Psi\ra$
of the  number operator $N(P)=\ad(P)a(P)$
in the quantum state $|\Psi\ra$ of the field. 
This does not fully characterize the state, 
but it will lead directly to considerations that do.

The annihilation operator (\ref{a(f)}) corresponding to a normalized
wavepacket $P$ is given by 
\beq
a(P)=\la P,\fphi\ra_{\Sigma_f},
\label{a(P)}
\eeq
where the
Klein-Gordon inner product (\ref{KGip}) is evaluated on the
``final" spacelike slice $\Sigma_f$. To evaluate the 
expectation value of $N(P)$ we use the field equation satisfied
by $\fphi$ to relate $N(P)$ to an observable on an 
earlier slice $\Sigma_i$ on which we know enough
about the quantum state.
Specifically, we assume there are no incoming 
excitations long after the black hole forms, and 
we assume that the state looks like the vacuum at 
very short distances (or high frequencies) 
as seen by observers falling across the event horizon. 
Hawking originally propagated the field through the
time-dependent collapsing  part of the metric and back out all the 
way to spatial infinity, where he assumed 
$|\Psi\ra$ to be the incoming 
vacuum at very high frequencies.  As pointed out by 
Unruh\cite{Unruh:ga} 
(see also \cite{Fredenhagen:1989kr,Jacobson:1993hn})
the result can be obtained without propagating
all the way back, but rather stopping on a spacelike
surface $\Sigma_i$ far to the past of $\Sigma_f$ 
but still after the formation of the black hole.
This is important since propagation back out to infinity
invokes arbitrarily high frequency modes whose 
behavior may not be given by the standard relativistic free
field theory.

If $\Sigma_i$ lies
far enough to the past of $\Sigma_f$,
the wavepacket $P$ propagated backwards
by the Klein-Gordon equation
breaks up into two distinct parts,
\beq
P=R+T.
\eeq
(See Fig. \ref{collapsefig}.)
$R$ is the ``reflected" part that scatters
from the black hole and returns to large radii, while 
$T$ is the ``transmitted" part that approaches 
the horizon. 
$R$ has support only at large radii, and $T$ has support
only in a very small region just outside the event horizon
where it oscillates very rapidly due to the backwards
gravitational blueshift.
Since both the wavepacket $P$ and the field operator $\fphi$ 
satisfy the Klein-Gordon equation, the Klein-Gordon inner 
product in (\ref{a(P)}) can be evaluated on $\Sigma_i$
instead of $\Sigma_f$ without changing 
$a(P)$. This yields
a corresponding decomposition for the annihilation
operator, 
\beq
a(P)=a(R)+a(T).
\eeq
Thus we have 
\beq
\la\Psi|N(P)|\Psi\ra=\la\Psi|(\ad(R)+\ad(T))(a(R)+a(T))|\Psi\ra.
\label{NRT}
\eeq

Now it follows from the stationarity of the black
hole metric that the Killing frequencies in a
solution of the KG equation are conserved.
Hence the 
wavepackets $R$ and $T$
both have the same, purely positive Killing, 
frequency components as $P$. As $R$ lies
far from the black hole in the nearly flat region,
this means that it has purely positive asymptotic
Minkowski frequencies, hence 
the operator $a(R)$ is a {\it bona fide} annihilation
operator---or rather $\la R,R\ra^{1/2}$ times the
annihilation operator---for incoming excitations.
Assuming that long after the black hole forms there 
are no such incoming excitations, we have $a(R)|\Psi\ra=0$.
Equation (\ref{NRT}) then becomes
\beq
\la\Psi|N(P)|\Psi\ra=\la\Psi|\ad(T)a(T)|\Psi\ra.
\label{NT}
\eeq
If $a(T)|\Psi\ra=0$ as well, then no $P$-particles
are emitted at all. The state with this property is 
called the ``Boulware vacuum". It is the state with no
positive Killing frequency excitations anywhere, including
at the horizon.

The Boulware vacuum 
does not follow from collapse however. 
The reason is that the wavepacket 
$T$ does not have purely
positive frequency with respect to the time of a free fall
observer crossing the horizon, and it is this latter frequency
that matches to the local Minkowski frequency in a
neighborhood of the horizon small compared with 
the radius of curvature of the spacetime.

More precisely,
consider a free-fall observer 
(i.e. a timelike geodesic $x(\t)$)
with proper time
$\t$ who falls across the horizon at $\t=0$ 
at a point where the slice $\Sigma_i$ meets
the horizon (see Fig. \ref{collapsefig}). 
For this observer the wavepacket $T$
has time dependence $T(\t)$ (i.e. $T(x(\t))$) that 
vanishes for $\t>0$ since the wavepacket has no
support behind the horizon. Such a function cannot
possibly have purely positive frequency components.
To see why, recall that if a function vanishes 
on a continuous arc in a domain of analyticity, then
it vanishes everywhere in that domain (since its
power series vanishes identically on the arc
and hence by analytic continuation everywhere).
Any positive frequency function
\beq
h(\t)=\int_0^\infty d\o \, e^{-i\o \t}\,  \tilde{h}(\o)
\label{posfreqint}
\eeq
is analytic in the lower half $\t$ plane, since the addition
of a negative imaginary part to $\t$ leaves the
integral convergent. The positive real $\t$ axis
is the limit of an arc in the lower half plane, 
hence if $h(\t)$ were to vanish for $\t>0$ it 
would necessarily vanish also for $\t<0$.
(Conversely, a function that
is analytic on the lower half-plane and
does not blow up exponentially
as $|\t|\rightarrow\infty$ must contain
only positive frequency components, since 
$\exp(-i\o \t)$ does blow up exponentially
as $|\t|\rightarrow\infty$ when $\o$ is negative.)

The wavepacket  $T$ can be decomposed 
into its positive and negative
frequency parts with respect to the free fall time $\t$,
\beq T = T^+ + T^-, \eeq
which yields the corresponding decomposition
of the annihilation operator
\bea a(T) &=& a(T^+) + a(T^-)\\
&=& a(T^+)-  \ad(\overline{T^-}).
\label{aT}\eea
Since $T^-$ has negative KG norm, 
Eqn. (\ref{ad(f)}) has been used in the last line to trade $a(T^-)$ 
for the {\it bona fide} creation operator
$ \ad(\overline{T^-})$.
The $\t$-dependence of $T$ consists
of very rapid oscillations for $\t<0$,
so the wavepackets
$T^+$ and $\overline{T^-}$ have very high energy 
in the free-fall frame.

A free fall observer crossing the horizon
long after the black hole forms would presumably
see the ground state of the field at short distances, 
that is, such an observer would see no very high
positive free-fall frequency excitations. The reason is that 
the collapse process occurs on the much longer time
scale of the Schwarzschild radius $r_s$, so the modes
with frequency much higher than $1/r_s$ should
remain in their ground state.
We therefore assume that the wavepackets $T^+$
and $\overline{T^-}$
are in their ground states,
\beq a(T^+)|\Psi\ra=0, \qquad a(\overline{T^-})|\Psi\ra=0.
\label{Uvacuum}
\eeq
For further discussion of this assumption see section
\ref{tpq}.

Using (\ref{aT})
the number expectation value (\ref{NT}) can be evaluated as
\bea
\la\Psi|N(P)|\Psi\ra&=&\la\Psi|a(\overline{T^-})\ad(\overline{T^-})|\Psi\ra\\
&=&\la\Psi|\bigl[a(\overline{T^-}),\ad(\overline{T^-})\bigr]|\Psi\ra\\
&=& \la\overline{T^-},\overline{T^-}\ra_{\Sigma_i}\\
&=& -\la{T}^-,{T}^-\ra_{\Sigma_i},
\label{NT-}
\eea
where (\ref{Uvacuum}) is used in the first and second lines,
(\ref{aadcomm}) is used in the third line, and 
(\ref{bracketrelns}) is used in the last step.
The problem has thus been reduced to the 
computation of the Klein-Gordon norm
of the negative frequency part of the 
transmitted wavepacket $T$. This
requires that we be more explicit about
the form of the wavepacket. 

\subsubsection{Norm of the negative frequency part \& thermal flux}
For definiteness
we consider a spherically symmetric vacuum
black hole in 3+1 dimensions, 
that is a {\it Schwarzschild} black hole.
The Schwarzschild line element can variously be expressed  as
\bea
ds^2&=&(1-\frac{r_s}{r})dt^2 - (1-\frac{r_s}{r})^{-1}dr^2-
r^2(d\theta^2 +\sin^2\theta\, d\fphi^2)\\
&=&(1-\frac{r_s}{r})(dt^2 - dr_*^2)-
r^2(d\theta^2 +\sin^2\theta\, d\fphi^2)\\
&=& (1-\frac{r_s}{r})du\, dv-
r^2(d\theta^2 +\sin^2\theta\, d\fphi^2).\label{null}
\eea
The first form is in ``Schwarzschild coordinates"
and $r_s=2GM$ is the Schwarzschild radius.
The second form uses the 
``tortoise coordinate" 
$r_*$, defined by $dr_*=dr/(1-r_s/r)$
or $r_*=r + r_s\ln(r/r_s -1)$, which goes to 
$-\infty$ at the horizon. The third form 
uses the retarded and advanced time
coordinates $u=t-r_*$ and $v=t+r_*$,
which are also called  
outgoing and ingoing 
null coordinates respectively.

A scalar field satisfiying the Klein-Gordon equation
$(\Box+m^2)\fphi=0$ can be decomposed into
spherical harmonics 
\beq
\fphi(t,r,\theta,\phi)=\sum_{lm}\frac{\fphi_{lm}(t,r)}{r}
Y_{lm}(\theta,\phi),
\eeq
where $\fphi_{lm}(t,r)$ satisfies the 1+1 dimensional
equation
\beq
(\partial_t^2 -\partial_{r_*}^2 + V_{lm})\fphi_{lm}=0
\eeq
with the effective potential
\beq
V_{lm}(r)=\Bigl(1-\frac{r_s}{r}\Bigr)\Bigl(\frac{r_s}{r^3} +\frac{l(l+1)}{r^2}+m^2\Bigr).
\eeq
As $r\rightarrow\infty$ the potential goes to $m^2$.
As $r\rightarrow r_s$, the factor $(r-r_s)$ approaches
zero exponentially as $\exp (r_*/r_s)$ with respect to 
$r_*$. Near the horizon $\fphi_{lm}(t,r_*)$ therefore satisfies
the massless wave equation, hence has the general form
$f(u) + g(v)$.

Since the wavepacket $P=\sum_{lm}P_{lm}(t,r)Y_{lm}(\theta,\phi)$
 is purely outgoing with support only 
at large radii at late times, near the 
horizon $P_{lm}(t,r)$ must be only a function of the `retarded
time $u=t-r_*$. That is, there can be no ingoing component.
Since the metric is static, i.e. invariant with respect 
to $t$-translations, we can decompose any solution into
components with a fixed $t$-frequency $\o$. 
A positive frequency outgoing mode
at infinity has $t$-dependence $\exp(-i\o t)$,  hence
its form near the horizon must be $\exp(-i\o u)$.

Consider now a late time outgoing positive 
frequency wavepacket $P$
that is narrowly peaked in Killing frequency $\o$.  Propagating
backwards in time, $T$ is the part of
the wavepacket that is squeezed up against the horizon,
and its $Y_{lm}$ component 
has the form $T_{lm}\sim \exp(-i\o u)$ for all $l,m$.
The coordinate $u$ diverges as the horizon is
approached. It is related to the proper time $\t$ of a 
free-fall observer crossing the horizon at $\t=0$ via 
$\t\simeq -\t_0\exp(-\k u)$,
where $\k=1/2r_s$ is the surface gravity of the black hole and the
constant $\t_0$ depends on the velocity of the free-fall
observer.\footnote{This can be obtained by noting from the
third form of the line element (\ref{null}) that along a
timelike line $(1-r_s/r)\dot{u}\dot{v}=1$,
where the dots represent the proper time derivative.
As the horizon is crossed $\dot{v}$ is finite, hence 
$\dot{u}\sim (r-r_s)^{-1}\sim e^{-r_*/r_s}= 
e^{(u-v)/2r_s}\sim e^{\k u}$.}
 Hence the $\t$-dependence of the wavepacket
along the free-fall worldline is
\beq
T \sim \exp\(i\frac{\o}{\k}\ln(-\t)\)
\label{Ttau}
\eeq
for $\t<0$, and it vanishes for $\t>0$.

To find the positive frequency part we use 
a method introduced by Unruh\cite{Unruh:db},
which exploits the fact that a function
analytic and bounded as $|\t|\rightarrow\infty$
in the lower half complex $\t$ plane 
has purely positive frequency 
(see the discussion after Eqn. (\ref{posfreqint})).
The positive frequency extension of $T(\t)$  
from $\t<0$ to $\t>0$ is thus obtained
by analytic continuation of $\ln (-\t)$
in the lower half complex 
$\t$-plane. This continuation is 
given by $\ln\t+i\pi$, 
provided the branch cut of $\ln\t$
is taken in the upper half-plane.
The positive frequency extension
of $T(\t)$ to $\t>0$ is therefore
obtained by replacing $\ln(-\t)$ 
with $\ln\t+ i\pi$ in (\ref{Ttau}), which yields 
$T(-\t)\exp(- \pi \o/\k)$ for $\t>0$.
Similarly, the negative frequency extension
of $\ln (-\t)$ is
given by $\ln\t-i\pi$, 
provided the branch cut of $\ln\t$
is taken instead in the lower half-plane.
The negative frequency extension
of $T(\t)$ to $\t>0$ is therefore
$T(-\t)\exp(+ \pi \o/\k)$. Knowing these
two extensions, we proceed as follows.

\begin{figure}[htb]
\vbox{ \vskip 8 pt
\centerline{\includegraphics[width=1.7in]{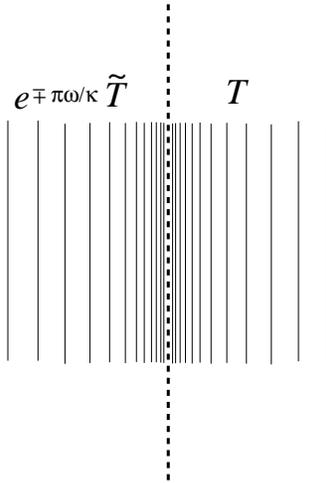}}
\caption{\label{partners}
\small Spacetime sketch of phase contours of
the transmitted wavepacket $T$
and its flipped version $\widetilde{T}$ on either
side of the horizon (dashed line). The upper and 
lower signs
in the exponent of the factor 
$\exp(\mp \pi\o/\k)$ yield the positive and negative 
free-fall frequency extensions of $T$. 
\smallskip}
} 
\end{figure}

Define a new wavepacket $\widetilde{T}$,
with support only inside the horizon,
by ``flipping"  the wavepacket $T(u)$ across
the horizon (see Fig. \ref{partners}). 
That is, $\widetilde{T}$ vanishes
outside the horizon and inside is
constant on the outgoing null lines,
with $\widetilde{T}(\t)=T(-\t)$ for $\t>0$.
The above argument shows that
the wavepackets
\bea
T^+&=&c_+(T+e^{-\pi\o/\k}\widetilde{T})\\
T^-&=&c_-(T+e^{+\pi\o/\k}\widetilde{T})
\label{Tpm}
\eea
have positive and negative free-fall frequency
respectively. The two constants 
$c_{\pm}$ can be chosen so that 
$T^++T^-$ agrees with $T$ outside the horizon
and vanishes (as does $T$) inside the horizon.
This yields $c_-=(1-e^{2\pi\o/\k})^{-1}$ and 
$c+/c_-=-e^{2\pi\o/\k}$.

Now $\la T,\widetilde{T}\ra
=0$ (since the two wavepackets do not overlap) 
and
$\la \widetilde{T},\widetilde{T}\ra
=-\la T,T\ra$ (since the flipped wavepacket
has the reverse $\t$-dependence), so using (\ref{Tpm})  
one finds
\beq
\la T^-,T^-\ra=\frac{\la T,T\ra}{1-e^{2\pi\o/\k}}.
\label{T-norm}
\eeq 
Inserting this in the expression (\ref{NT-}) 
for the number operator expectation
value yields Hawking's result,
\beq
\la\Psi|N(P)|\Psi\ra =\frac{\la T,T\ra}{e^{2\pi\o/\k}-1}.
\label{NH}
\eeq

The number expectation value (\ref{NH}) corresponds
to the result for a thermal state 
at the Hakwing 
temperature $T_H=\hbar \kappa/2\pi$, multiplied by 
the so-called ``greybody factor" 
\beq
\Gamma=\la T,T\ra.
\label{greybody} 
\eeq
This factor
is the probability for an excitation described
by the the wavepacket $P$ to pile up just outside the 
event horizon when propagated backwards in time,
rather than being scattered back out to infinity.
Equivalently, $\Gamma$ is the probability
for the excitation with wavepacket
$P$ to fall across 
the horizon when sent in forwards in time. 
This means that the black hole would be in detailed
balance with a thermal bath at the Hawking temperature.
Yet another interpretation of $\Gamma$ is the probability
for an excitation originating close to the horizon with 
normalized wavepacket $T/\la T,T\ra^{1/2}$ to 
escape to infinity rather than scattering back and falling
into the black hole.

\subsubsection{The quantum state} 

The local free-fall vacuum condition (\ref{Uvacuum})
can be used to find the quantum state of the near 
horizon modes. We can do that here just ``pair by pair",
since the field is noninteracting so the state is
the tensor product of the states for each outside/inside 
pair.

Using (\ref{Tpm}) the vacuum conditions
become
\bea
a(T^+)|\Psi\ra&\propto& \[a(T)-e^{-\pi\o/\k}\ad(\tilde{T}^*)\]|\Psi\ra=0
\label{noT+}\\
a(\overline{T^-})|\Psi\ra&\propto &\[-\ad(T)+e^{+\pi\o/\k}a(\tilde{T}^*)\]|\Psi\ra=0
\label{noT-}
\eea
(here we use $*$ instead of ``bar" 
for complex conjugation of $\widetilde{T}$ for typographical reasons).
Note that since $\overline{T}$ and $\widetilde{T}$  
have negative norm (as explained before (\ref{T-norm})), we have replaced the corresponding annihilation operators by minus the creation operators of their conjugates. These
equations define what is called the {\it Unruh vacuum} $|U\ra$
for these wavepacket modes.

Now let $|B\ra$ denote the quantum state 
of the $T$ and $\widetilde{T}^*$ modes 
such that
\bea 
a(T)|B\ra&=&0\\
a(\widetilde{T}^*)|B\ra&=&0.
\eea
This state is called the {\it Boulware vacuum}
for these modes.
In analogy with (\ref{gnd}) and (\ref{squeezedgnd})
the vacuum conditions (\ref{noT+},\ref{noT-})
imply that
the Unruh and Boulware vacua are related by 
\beq
|U\ra\propto
\exp\[e^{-\pi\o/\k}\ad(\hat{T})\ad(\hat{\widetilde{T}^*}) \]
|B\ra
\eeq
where the hats denote the corresponding normalized
wavepackets. 

The Unruh vacuum is thus a two-mode squeezed
state, analogous to the one (\ref{squeezedvac}) 
found at each 
wavevector $\bf k$ in the case of cosmological 
pair creation. There each pair had zero total momentum
since the background was space translation invariant. In the
present case, each pair has zero total Killing energy,
since the background is time translation invariant. 
The mode $\widetilde{T}$ has the same positive Killing 
frequency as $T$ has (because the Killing 
flow is symmetric under the flipping across the horizon
operation that defines $\widetilde{T}$), 
however its conjugate has negative
Killing frequency, and therefore negative Killing energy.

The Unruh vacuum is a pure state, but because of
its entangled structure it becomes mixed 
when restricted to the exterior. To find that mixed state
we expand the exponential in a series.
Denoting by $|n_{R,L}\ra$ the level-$n$ excitations 
of the modes $T$ and $\widetilde{T}^*$ respectively, we have
\beq
|U\ra\propto  \sum_n e^{-n\pi\o/\k}|n_L\ra|n_R\ra.
\label{U}
\eeq
The reduced density matrix is thus 
\beq
Tr_L|\\U\ra\la U| \propto \sum_n e^{-2n\pi\o/\k}|n_R\ra\la n_R|,
\label{Uright}
\eeq
a thermal canonical ensemble at the Hawking temperature.

The essence of the Hawking effect is 
the correlated structure of the local vacuum state at short distances near the horizon and its thermal character outside the horizon. Given this, the outgoing flux at infinity is just a consequence of the propagation of a fraction $\G$ (\ref{greybody})
of each outgoing wavepacket from the horizon to infinity.

\subsection{Remarks}

In this subsection we make a large number of brief remarks
about related topics that we have no time
or space to go into deeply. Where no references are given see
the sources listed at the end of the Introduction.

\subsubsection{Local temperature} 
\label{localtemp}

The Hawking temperature refers to the Killing energy,
or, since the Schwarzschild Killing vector is normalized at 
infinity, to the energy defined by a static observer
at infinity. A static observer at finite
radius will perceive the thermal state to have the 
blueshifted temperature
$T_{loc}=T_H/|\xi|$, where $|\xi|$ is the local norm of the 
Schwarzschild time
translation Killing vector. At infinity this is 
just the Hawking temperature, whereas 
it diverges as the horizon is approached.  This divergence
is due to the infinite acceleration of the static observer at the horizon
and it occurs even for an accelerated observer
in the Minkowski vacuum of flat spacetime (see section
\ref{Unruheffect} below).
A freely falling observer sees nothing divergent.

\subsubsection{Equilibrium state: Hartle-Hawking vacuum}

A black hole will be in equilibrium with an incoming thermal flux
at the Hawking temperature. The state that includes this incoming flux is called the Hartle-Hawking vacuum.

\subsubsection{Stimulated emission}
\label{stim}

Suppose that the field is not
in the free-fall vacuum at the horizon 
(\ref{Uvacuum}), but rather
that there are $n$ excitations in the mode $T^+$,
so that $\ad(T^+)a(T^+)|\Psi\ra=n\la T^+,T^+\ra|\Psi\ra$.
Then instead of (\ref{NT-}) the expectation value of the 
number operator will be 
\beq
\la\Psi|N(P)|\Psi\ra=n\la T,T\ra+(n+1)\la\overline{T^-},\overline{T^-}\ra.
\label{Nstim}
\eeq
That is, if $n$ quanta are present to begin with 
in the $T^+$ mode, the observer at infinity will observe
in the $P$ mode $n+1$ times the usual number of 
Hawking quanta, in addition to $n$ times the greybody factor
(\ref{greybody}).
To produce a state in which the $T^+$ mode is occupied 
in standard physics
one would have to send in particles of enormous energy just
before the black hole formed\cite{Wald:ka}. As explained
in section \ref{latestim}, however, 
trans-Planckian considerations could in principle
allow stimulated emission
at times long after the black
hole formed. (This has nothing to do with the standard
late time stimulated emission of the super-radiant 
modes of a rotating black 
hole\cite{Bekenstein:mv,Panangaden:pc}.)

\subsubsection{Unruh effect}
\label{Unruheffect}

The argument given above for the structure of the vacuum
near a black hole horizon
applies equally well to the Minkowski vacuum near an
acceleration horizon in flat spacetime, 
where it is known as the Unruh effect. 
From a logical point of view it might be better
to introduce the Unruh effect first, and then 
export it to the neighborhood of a black hole horizon 
to infer the Hawking effect. However, 
I chose here to go in the other direction

In the Unruh effect
the boost Killing field $\xi_B=x\partial_t+t\partial_x$ 
(which generates hyperbolic rotations) plays the role of 
the Schwarzschild time translation, 
and the corresponding ``temperature" is $\hbar/2\pi$. 
The Minkowski vacuum is the 
analog of the Hartle-Hawking equilibrium state, rather than
the Unruh evaporating state. 
A uniformly accelerated observer following
a hyperbolic orbit of the Killing field will perceive the 
Minkowski vacuum
as a thermal state with temperature $\hbar/2\pi|\xi_B|$.
The norm $|\xi_B|$ is just $(x^2-t^2)^{1/2}$, which is also
the inverse of the acceleration of the orbit, hence the local
temperature is the {\it Unruh temperature} $T_U=\hbar a/2\pi$.
As $(x^2-t^2)^{1/2}\rightarrow\infty$ this temperature is redshifted to zero,
so a Killing observer at infinity sees only the zero temperature
vacuum. As the acceleration horizon $x=\pm t$ is approached
a Killing observer sees a diverging temperature. The
same temperature divergence is seen by a static observer
approaching the horizon of a black hole in the Unruh or 
Hartle-Hawking states ({\it cf.} section \ref{localtemp} above).

\subsubsection{Rotating black hole}

A small portion of the event horizon of a rotating black hole
is indistinguishable from that of a Schwarzschild black hole,
so the Hawking effect carries over to that case as well.
The frequency $\o$ in (\ref{NH})
should be replaced by the frequency with 
respect to the horizon generating Killing field $\partial_t +\O_H
\partial_\phi$, where $\partial_t$ and $\partial_\phi$ are the asymptotic time translation 
and rotation Killing vectors, and $\O_H$ is the angular velocity of the horizon. Thus $\o$ is replaced by $\o -m\O_H$, where $m$ denotes
the angular momentum. In effect there is a chemical potential
$m\O_H$. See  e.g. \cite{Frolov:jh} for a discussion of the
quantization of the ``super-radiant" modes with $\o -m\O_H<0$.

\subsubsection{de Sitter space}

The reasoning in the black hole case applies {\it mutatis mutandis}
to de Sitter spacetime, where an observer is surrounded by a 
horizon that is locally indistinguishable from a black hole horizon.
This leads to the temperature of deSitter spacetime\cite{GH,AndydS}.

\subsubsection{Higher spin fields}

The Hawking effect occurs also for
higher spin fields, the only difference being
(1) the greybody factors are different, and (2)
for half-integer spin fields the Fermi distribution rather than
the Bose distribution arises for the Hawking emission.

\subsubsection{Interacting fields}

Our discussion here exploited the free field equation of motion,
but the Hawking effect occurs for interacting fields as well.
The essence, as in the free field case, is
the Unruh effect, the interacting version of which can easily
be established with the help of a Euclidean functional integral
representation of the Minkowski vacuum\cite{Unruh:1983ac}.
(This result was found a decade earlier, 
at about the same time as the original Unruh effect,
via a theorem\cite{BisoWich} in the context of axiomatic quantum field theory, although the interpretation in terms of the thermal observations of uniformly accelerated observers was not noted 
until later\cite{Sewell}.) The direct analog for the Hawking effect 
involves a Euclidean functional integral expression for the
interacting  Hartle-Hawking equilibrium state 
(for an introduction see \cite{TJnote,TJmicro} and references therein). 

More directly, in an asymptotically free theory one can 
presumably use the free field analysis to discover the
structure of the vacuum near the horizon as in the
free field case. The propagation of the field from that
point on will involve the interactions. If the Hawking temperature
is much higher than the scale $\L$ of asymptotic freedom then
free particles will stream away from the black hole and subsequently 
be ``dressed" by the interactions and fragment
into asymptotic states\cite{MacGibbon:zk, MacGibbon:tj}. 
If the Hawking temperature is much
lower than $\L$ then it is not so clear (to me at least) how
to determine what is emitted.

\subsubsection{Stress-energy tensor}

The Unruh and Hartle-Hawking states are ``regular"
on the horizon, i.e. they look like the Minkowski
vacuum at  short distances. 
(Recall that it is precisely the local vacuum property
(\ref{Uvacuum}) or (\ref{noT+},\ref{noT-}) 
that determines the thermal
state of the outgoing modes at infinity.)
Hence the mean value
of the stress energy tensor is finite. The Boulware state
$|B\ra$ referred to above is obtained by removing
the $T$, $\widetilde{T}^*$ pair excitations. This produces a
state with a negative mean energy density that diverges
as the horizon is approached. In fact, 
if even just {\it one} Hawking quantum is removed from 
the Unruh state $|U\ra$ to obtain the Boulware state
for that mode, a negative energy density divergence
will be produced at the horizon. This can be viewed as the
result of the infinite blueshift of the negative energy ``hole".
That is quite odd in the
context of flat space, since the Minkowski vacuum should
be the lowest energy state, hence any other state should
have higher energy. The explanation is that there
is a positive energy density divergence {\it on}
the horizon that more than compensates the negative
energy off the horizon\cite{RPpuzzle}.

\subsubsection{Back-reaction}

As previously noted the Unruh state corresponds to 
an entangled state of positive and negative Killing energy
excitations. As the positive energy excitations escape to infinity,
there must be a corresponding negative energy flux into the black hole. Studies of the mean value of the stress tensor confirm this.
Turning on the gravitational dynamics, this would
lead to a mass loss for the black hole via the Einstein equation.
The backreaction driven by the mean value is called the
``semi-classical" evolution. There are quantum fluctuations
about this mean evolution on a time and length scale of the
Schwarzschild radius, unless a large number of matter fields is invoked to justify a large $N$ limit that suppresses the quantum
fluctuations.

\subsubsection{Statistical entropy} 

The entangled structure (\ref{U}) of the 
Unruh state leads to a mixed state (\ref{Uright})
when observations are restricted to the region
outside the horizon. The ``entanglement entropy" 
$-Tr \rho\ln\rho$ of this state is the same as the
thermal entropy of the canonical ensemble (\ref{Uright}). 
Summing over all modes
this entropy diverges due to the infinite density of modes.
To characterize the divergence one can use the thermodynamic
entropy density $s\propto T^3$ of a bath of radiation
at temperature $T$.
The local temperature 
measured by a static observer ({\it cf.} section \ref{localtemp})
is given by  
\beq
T_{loc}=T_H/|\xi|\simeq T_H/\kappa \ell =1/2\pi \ell,
\eeq
where $\ell$ is the proper distance to the horizon on 
a surface of constant $t$, and the relation 
$\kappa = (d|\xi|/d\ell)_H$ has been used. Thus the entropy diverges like
\beq
S=\int s\;  dv \sim \int T_{loc}^3 \; d\ell d^2A\sim A/\ell_c^2,
\eeq
where $\ell_c$ is a cutoff length above the horizon.\footnote{
Sorkin\cite{SorkinGR10} introduced the notion of 
black hole entanglement entropy, and with collaborators\cite{Bombelli:1986rw} 
computed it in the presence of a regulator.  A mode-by-mode
version of the thermal entropy calculation 
was first done by 't Hooft\cite{'tHooft:1984re}, who called it the 
``brick wall model" because of a Dirichlet boundary condition  
applied at $\ell_c$.} 

What is the meaning of this entropy? It seems clear on the
one hand that it must be included in the black hole entropy,
but on the other hand it must somehow be meaningfully cut
off. It is natural to try to understand the scaling of the black hole entropy with area in this way, but this is only the contribution from one quantum field, and there is also the classical contribution from the gravitational field itself that emerges from the partition function for quantum gravity\cite{GHpf}. The apparent dependence on the number of fields is perhaps removed by the corresponding renormalization of Newton's constant\cite{G}. (The last reference in \cite{G} is a review.)
However this is regularization dependent and hence difficult to interpret physically, and moreover at least in dimensional regularization vector fields and some non-minimally coupled scalars
contribute {\it negatively} to renormalizing $G$,
which does not seem to match the entanglement entropy.
There has been much work in this area but it remains
to be fully understood.

\subsubsection{Information loss}
\label{infoloss}

Two types of potential information loss occur in black hole
physics. First, when something falls into a black hole any information it carries is apparently lost to the outside world.
Second, when a black hole radiates Hawking quanta, each
radiated quantum is entangled with a partner lying inside the
horizon, as Eqn. (\ref{U}) shows. As long as the black hole
does not evaporate completely, all information remains available on a spacelike surface that crosses the horizon and enters the black hole. If on the other hand the black hole evaporates completely, then no single spacelike surface stretching to 
infinity and filling ``all space" can capture all
information, according to the semi-classical analysis of the 
Hawking effect.
Rather a disconnected surface behind the
horizon must be included. The information on this disconnected surface flows into the strong curvature region at the singularity,
where its fate is not yet understood.

This situation has generated much discussion. 
Some researchers (myself included) 
see the information loss
to the outside world not as a sign of the breakdown of
quantum mechanics but just as a consequence of the mutability of spatial topology in quantum gravity. When a black hole 
is about to evaporate completely it looks very small to the
world outside the horizon. However this outside smallness has
absolutely nothing to do with the size of the region inside
available for storing information. Let us look into this
a bit more.

Consider for example
the spacelike singularity at $r=0$ inside the Schwarzschild black 
hole. The metric in Eddington-Finkelstein coordinates
is
\beq
ds^2 = (1-r_s/r)dv^2 -2 dv dr - r^2(d\theta^2 + \sin^2\theta d\phi^2),
\eeq
where $v=t+r_*$
is the advanced time 
coordinate defined below (\ref{null}). Inside the horizon,
where $r<r_s$, a line of constant $r,\theta,\phi$ is spacelike,
and has a proper length $L(r, \D v)=(r_s/r-1)^{1/2}\D v$.
This goes to infinity as the singularity is approached, 
{\it for any interval of advanced time} $\D v$. Hence there
is no dearth of space inside. On the other hand, the transverse
angular dimensions go to zero size, and also we do not know
how to describe the spacetime too near to the singularity.
Therefore let's stop at the radius where the curvature $\sim r_s/r^3$
is equal to the Planck curvature, i.e at $r\sim r_s^{1/3}$
in Planck units. Then then proper length goes as 
$L(r_s^{1/3}, \D v)\sim r_s^{1/3}\D v$. For a solar mass black
hole we have $r_s\sim 1\,  {\rm km} \sim 10^{38}$ in Planck
units, so $r_s^{1/3}\sim 10^{13}$. That means that 
for an external advanced time of $\D v=1$ second, the proper 
length inside is one million light years. After a day or so, 
the length is the size of the universe, and so on over the Hawking
lifetime $r_s^3\sim 10^{114}$. 

What happens to the future of this {\it spacelike} 
$r=r_s^{1/3}$ cylinder inside the black hole  
is governed by quantum gravity. 
We don't know what form the evolution takes. It is 
conceivable that time just stops running, like a frozen engine,
producing a boundary of spacetime. It seems much more likely however
that spacetime persists beyond, either into ``quantum foam" or into 
a plump baby universe 
(see e.g. \cite{Dymnikova:2003vt} and references therein) 
or universes\cite{Barrabes:1995nk}.
In any of these scenarios, the tiny outside size of a black hole 
at the end stage of Hawking evaporation 
is no indication of the information carrying capacity of the 
interior. 

Others believe that
the validity of quantum mechanics requires that all information 
winds up available on the exterior slice after the evaporation. 
Results from string theory are often invoked to support this
viewpoint. (For a critique of these arguments see \cite{TJBHentropy}.)
If true, it would require some breakdown of the semi-classical description 
where none seems to be otherwise called for.
The role of ultra high frequencies in the Hawking effect
is often brought up in this context, however they are
irrelevant since the derivation of the correlated structure of
the vacuum and the Hawking effect does not need
to access those frequencies. For more on this see the
discussion of the trans-Planckian question in section \ref{tpq}.

\subsubsection{Role of the black hole collapse}
\label{collapse}

The Hawking flux continues in a steady state long after the collapse
that forms a black hole, which suggests that the collapse phase 
has nothing to do with the Hawking effect. Indeed the derivation
given above makes use of only the free-fall vacuum conditions
near the horizon, and the collapse phase plays no role.
On the other hand if, as Hawking originally did, we follow the
positive Killing frequency 
wavepacket $T$ all the way backwards in time, it would go through the 
collapse phase and back out to infinity where Killing frequency and 
free-fall frequency coincide. Were it not for the time dependence
of the background during the collapse, there would be no change of
the Killing frequency, so there would be no negative frequency 
part of the ingoing wavepacket and there would be no particle
creation. As discussed in the following section this propagation
through the collapse phase invokes  ultra high frequency
field modes, and therefore may  not even be physically
relevant. However it seems clear is that whatever physics
delivers the outgoing vacuum near the horizon, it must involve
some violation of the time translation symmetry of the
classical black hole background, even if it does not 
involve the collapse phase. In the lattice model of 
section \ref{lattice} the violation arises from microscopic time
dependence of the lattice spacing. In quantum gravity
it may come just from the quantum gravitational fluctuations
(see section \ref{latticetimedep}).

\sect{The trans-Planckian question}
\label{tpq}

A deep question arises on account of the infinite redshift
at the black hole horizon: do the outgoing
modes that carry the Hawking radiation really 
emerge from a  reservoir of modes with frequency
arbitrarily far beyond the Planck frequency
just outside the horizon, or is there another possibility?
Does the existence or properties of the Hawking effect
depend on the existence of such a {\it trans-Planckian
reservoir}?

The reasoning leading to the expression (\ref{U})
for the Unruh state in terms of positive Killing frequency
modes is largely shielded from this question. The essential
input is the free-fall vacuum conditions (\ref{Uvacuum}),
which can be applied at any length or time scale much shorter
than the Schwarzschild radius and inverse surface gravity
(which are roughly the same unless the black hole is near 
extremally rotating or charged). There is no need to 
appeal to Planckian or trans-Planckian frequencies. 

From this persepctive it is clear that, as far as the
derivation of the Hawking effect
is concerned, the only question  is whether or not the
free-fall vacuum in fact arises at short distances near the
horizon from the initial conditions
before collapse. As mentioned earlier, the modes with 
frequencies much higher than the inverse of the collapse 
time scale would be expected to remain unexcited.
Nevertheless, in standard relativistic field theory these
modes arise from trans-Planckian modes. 

Consider an outgoing wavepacket near the horizon 
and peaked around frequency
$\o_1^{\rm ff}$ as measured by a free-fall observer crossing the
horizon at the advanced time $v_1$. At an earlier time
$v_2=v_1-\D v$, 
the wavepacket would be blueshifted 
and squeezed closer to the horizon, 
with exponentially higher free-fall frequency, 
\beq
{\o_2^{\rm ff}}/{\o_1^{\rm ff}}\sim e^{\k \D v} = e^{\D v/2 R_s}.
\eeq
For a solar mass black hole and $\D v=2$ seconds, the ratio 
is $\exp(10^5)$.

To {\it predict} the state of the positive free-fall frequency modes
$T^+$ and $\overline{T^-}$ from the initial state thus seems to require trans-Planckian physics. This is a breakdown of the 
usual separation of scales invoked in the application of effective field theory and it leaves some room for doubt\cite{Jacobson:1991gr,Jacobson:1993hn,Helfer:2003va} 
about the existence of the Hawking effect. 

While the physical arguments for the Hawking effect 
do seem quite plausible, the trans-Planckian question 
is nevertheless pressing. After all, there are reasons 
to suspect that the trans-Planckian modes do not even 
{\it exist}. They imply an infinite contribution to 
black hole entanglement entropy from quantum fields, and they produce
other divergences in quantum field theory that are not desirable in a fundamental theory. 

The trans-Planckian question is really two-fold:
\ben
\item Is the Hawking effect {\it universal}, i.e. insensitive to short distance physics, or 
at least can it be reliably derived in a quantum gravity theory with 
acceptable short distance behavior?
\item If there is no trans-Planckian reservoir, from where do the outgoing black hole modes arise?
\een

\subsection{String theory viewpoint}
String theory has made impressive progress towards answering the first question, at least for some special black holes. In particular\cite{Horowitz:1996rn,Peet:1997es}, some near-extremal black holes become well-understood D-brane configurations in the weak coupling limit, and supersymmetry links the weak coupling to strong coupling results. Thus the Hawking effect can be reliably analyzed in a full quantum gravity theory. 
By a rather remarkable and unexpected correspondence the computations yield agreement with the semi-classical predictions, at least in the long wavelength limit.
Moreover, the D-brane entropy is understood in terms of the counting of microstates, and agrees with the corresponding black hole entropy at strong coupling, just as the supersymmetry reasoning says it should. From yet another angle, the AdS/CFT duality in string theory offers other support\cite{magoo}.  There the Hawking effect and black hole entropy are interpreted in terms of a thermal state of the CFT (a conformally invariant super-Yang-Mills
theory). However, neither of these approaches from string theory has so far been exploited to address the origin of the outgoing modes, since a local spacetime picture of the black hole horizon is lacking. This seems to be a question worth pursuing.

\subsection{Condensed matter analogy}
Condensed matter physics provides an analogy for effective field theory with a fundamental cutoff, hence it can be used to explore
the consequences of a missing trans-Planckian reservoir.
(For a review of these ideas see \cite{river}, and for a very brief
summary see \cite{indyhawk}.) 
The first such black hole analog was Unruh's sonic black hole, which consists of a fluid with an inhomogeneous flow exceeding
the speed of sound at a sonic horizon. A molecular fluid does not
support wavelengths shorter than the intermolecular spacing, hence the sonic horizon has no ``trans-molecular" reservoir
of outgoing modes. 
Unruh found that nevertheless outgoing modes are produced, 
by a process of  ``mode conversion" from ingoing to outgoing modes. 
This phenomenon comes about already because
of the alteration of the dispersion relation for the sound waves.
It has been studied in various field theoretic models, however
none are fully satisfactory since the unphysical short distance
behavior of the field is always eventually called into play. The model
that most closely mirrors the fluid analogy is 
a falling lattice\cite{lattice} which has sensible short distance physics. 
The mode conversion on the lattice involves what is 
known as a Bloch oscillation in the condensed matter 
context. Here I will briefly explain how  it works. 

\subsection{Hawking effect on a falling lattice}
\label{lattice}

The model is 2d field theory on a lattice of points falling 
freely into a black hole. We begin with the line element in
Gaussian normal coordinates,
\beq
ds^2 = dt^2 -a^2(t,z) \, dz^2.
\eeq
A line of constant $z$ is an infalling geodesic, at rest at infinity, 
and the ``local scale factor" $a(t,z)$ satisfies
\bea
a(t,z\rightarrow\infty)&=&1\\
a(0,z)&=&1\\
a(t,z_H(t))&\sim &\k t \qquad {\rm for}\; \k t\gsim 1,
\eea
expressing the facts that the metric is asymptotically
flat, the coordinate $z$ measures proper distance on the
$t=0$ time slice, and at the horizon $z_H(t)$ the time scale
for variations of the local scale factor is the surface gravity
$\k$. The specific form of $a(t,z)$ is not required for the
present discussion. (For details see \cite{lattice}.)
A scalar field on this spacetime is governed by the action
\bea
S&=&\frac{1}{2}\int d^2x\; \sqrt{-g}g^{\mu\nu}\, \partial_\mu\fphi\, \partial_\nu\fphi \\
&=&\frac{1}{2}\int dt dz\left[a(z,t)(\partial_t\fphi)^2-\frac{1}{a(t,z)}(\partial_z\fphi)^2\right].
\eea

Now we discretize the $z$ coordinate with spacing $\d$:
\beq
z\rightarrow z_m = m\d, \qquad \partial_z\fphi \rightarrow D\fphi=\frac{\fphi_{m+1}-\fphi_m}{\d},
\eeq
where $\fphi_m=\fphi(z_m)$. 
The proper distance between the lattice points $z_m$ and $z_{m+1}$ 
on a constant $t$ slice is approximately $a(t,z_m)\d$. At $t=0$ this is just
$\d$ everywhere, that is the points start out equidistant. However, since they are
on free fall trajectories at different distances from the horizon, they spread out 
as time goes on. In particular the lattice spacing at the horizon grows with time
like $\sim \k t$.

\begin{figure}[htb]
\vbox{ \vskip 8 pt
\centerline{\includegraphics[width=2.7in]{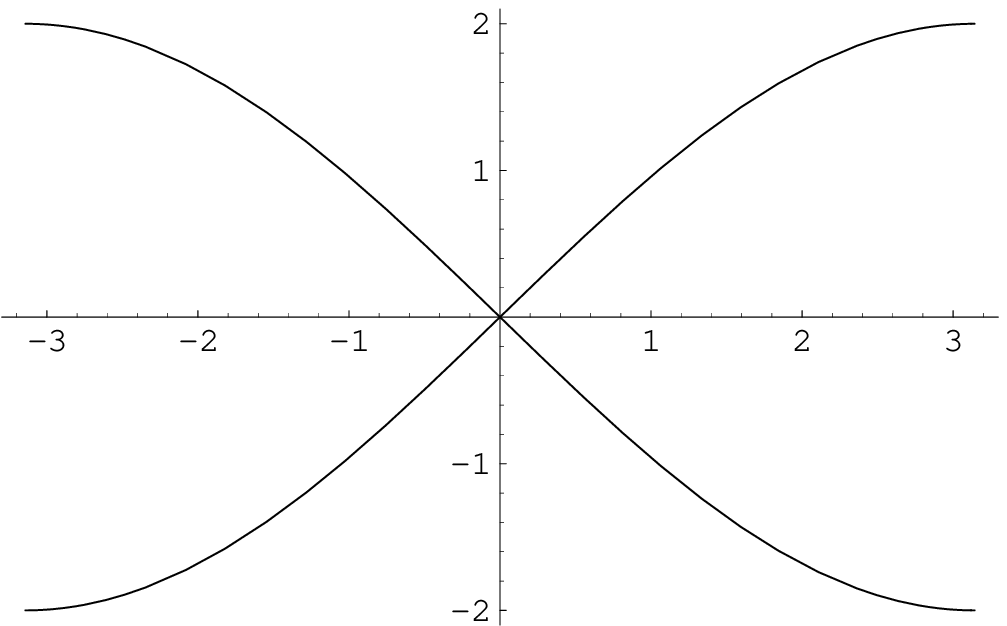}}
\caption{\label{lattdr}
\small Dispersion 
relation $\o\d=\pm 2\sin(k\d/2)$ 
plotted vs. $k\d$.  Wavevectors differing
by $2\pi/\d$ are equivalent.
Only the Brillouin zone 
$|k|\le\pi/\d$ is shown.
\smallskip}
} 
\end{figure}

The discretized action is
\beq
S_{\rm lattice}=\frac{1}{2}\int dt \sum_m\left[a_m(t)(\partial_t\fphi_m)^2-\frac{2(D\fphi_m(t))^2}{a_{m+1}(t)+a_m(t)}\right].
\eeq
The discrete field equations produce the dispersion relation
\beq
\o^{\rm ff}(k)=\pm \frac{2}{a(z,t)\d}\sin(k\d/2)
\eeq
for a mode of the form $\exp(-i\o^{\rm ff}t+ikz)$,
provided $\partial_t a \ll \o^{\rm ff}$ and 
$\partial_z a\ll k$. (See Fig. \ref{lattdr}.)
For small wave numbers the lattice dispersion agrees with the 
continuum, however it is periodic in translation of $k$ by 
$2\pi/\d$. At the wavenumber $k=\pi/\d$ there is a maximum 
frequency $2/a\d$ and vanishing group velocity $d\o/dk$ . 
Beyond that wavenumber the group velocity reverses,
and  $k$ is equivalent to $k-2\pi/\d$ which  lies in the Brillouin zone 
$|k|\le\pi/\d$.

\begin{figure}[htb]
\vbox{ \vskip 8 pt
\centerline{\includegraphics[width=6in]{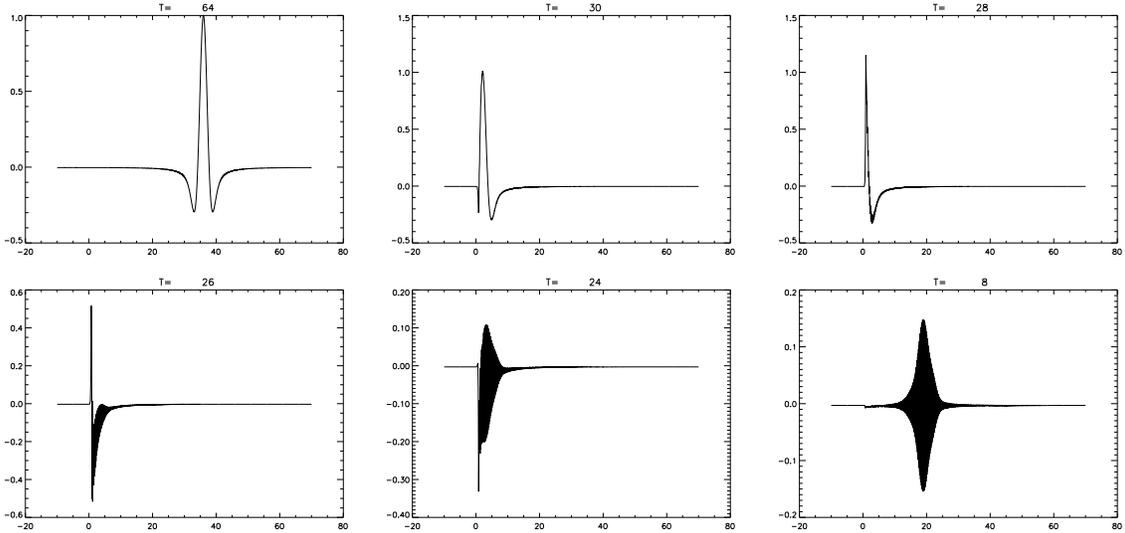}}
\caption{\label{tsequence}
\small A typical wavepacket evolution on the lattice.
The oscillations of the incoming wavepacket are too dense 
to resolve in the plots. 
\smallskip}
} 
\end{figure}

On the lattice the trans-Planckian redshift cannot take place,
because of the lattice cutoff. Hence  the outgoing modes---provided 
they exist---must come from ingoing modes. Both WKB 
``eikonal" trajectory and numerical evolution of the discrete wave equation 
confirm that indeed this occurs.
The behavior of a typical wavepacket throughout
the process of bouncing off the horizon is illustrated
in Fig. \ref{tsequence}. 
The real part of the
wavepacket is plotted vs. the static coordinate at several
different times. Following backwards in time,
the packet starts to squeeze up against the
horizon and then a trailing dip freezes and develops oscillations that grow
until they balloon out, forming into a compact high frequency wavepacket
that propagates neatly away from the horizon backwards in time. 

The mode conversion can be understood as follows, following a
wavepacket peaked around a long wavelength $\l\gg\d$
backwards in time. The wavepacket blueshifts 
as it approaches the horizon, eventually enough for the lattice structure and therefore the curvature of the dispersion relation to be felt. At this point its group velocity begins to drop. In the WKB calculation the wavepacket motion reverses direction 
at a turning point outside the horizon.
This occurs before its group velocity in the falling lattice frame is
negative, so it is falling in at that stage because its outward velocity is not great enough to overcome the infalling of the lattice.
 
As the wavepapcket  continues backwards in time now away 
from the black hole its wavevector continues to grow until it goes past the 
edge of the Brillouin zone, thus becoming an ingoing
mode also in the lattice frame. This reversal of group velocity
is precisely what happens
in a ``Bloch oscillation" when a quantum particle in a periodic potential
is accelerated. As the turnaround is occuring,
another equally important effect is that the time-dependence
of the underlying lattice is felt. Thus, unlike in the continuum 
limit of this stationary background,  {\it the Killing frequency of the wavepacket is
no longer conserved}. 

Following the wavepacket all the way backwards in time out
to the asymptotic region, it winds up with a short
wavelength of order $\d$ and a large frequency  of order
$1/\d$. This frequency shift is absolutely critical to the
existence of the outgoing modes, since an ingoing low
frequency mode would simply fall across the horizon. Only
the ``exotic" modes with sufficiently high frequency will
undergo the mode conversion process. 

The Hawking flux is determined by the negative
frequency part of the ingoing wavepacket. The eikonal approximation
just described does not capture the negative frequency mixing
which occurs during the turnaround at the horizon. Just as in the
continuum, the wavepacket squeezed against the horizon has
both positive and negative free-fall frequency components. 
As these components propagate backwards in time away from the horizon,
their frequency slowly shifts, but their relative amplitude remains
fixed. Hence the norm of the negative freqency part of the 
incoming wavepacket  turns out to be just what
the continuum Hawking effect indicates, with small lattice corrections.
Put differently, the infalling vacuum is adiabatically
modified by the underlying microscopic time dependence of the 
lattice in such a way that the Unruh conditions (\ref{Uvacuum}) on the 
state of the outgoing modes at the horizon are satisfied. 

\subsection{Remarks}

\subsubsection{Finite Entanglement entropy}
\begin{figure}[thb]
\vbox{ \vskip 8 pt
\centerline{\includegraphics[angle=-90,width=2.7in]{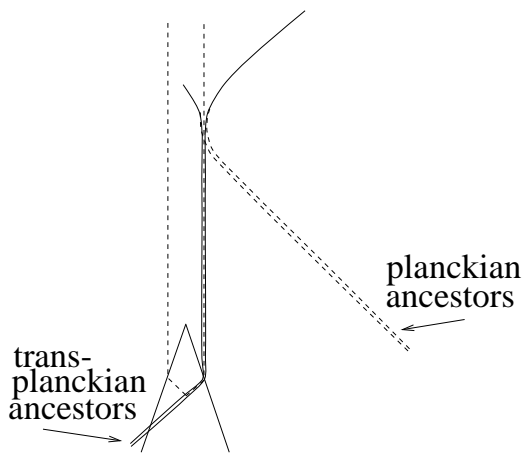}}
\caption{\label{ancestors}
\small The ancestors of a Hawking quantum 
and its negative energy partner. In standard relativistic field theory
the ancestors are trans-Planckian and pass through the
collapsing matter at the moment of horizon formation. On the
lattice the ancestors ar Planckian and propagate in towards 
the black hole at late times.
\smallskip}
} 
\end{figure}

Since the lattice has a short distance cutoff 
the entanglement entropy between the modes 
just inside and outside of the horizon is finite at any time.
As illustrated in Figure \ref{ancestors},
any given entangled pair of vacuum modes 
began in the past outside the
horizon, propagated towards the horizon where it was
``split", and then separated, with one half falling in
and the other converted into an outgoing mode. 
As time goes on, new pairs continually propagate in
and maintain a constant entanglement entropy.

\subsubsection{Stimulated emission of Hawking radiation
at late times}\label{latestim}
As  discussed in section  \ref{stim},
if the Unruh vacuum conditions (\ref{Uvacuum}) do not 
hold at the horizon then stimulated emission of Hawking 
radiation will occur. In the falling lattice model, these 
horizon modes arise from ingoing modes long after the black
hole formed as shown in Figure \ref{ancestors}. 
Thus it is possible to stimulate the emission of 
Hawking radiation by sending in radiation at late times, in contrast to the
usual continuum case of a static black hole.
This seems a generic feature of theories with a cutoff,
for which the outgoing modes must arise from modes
that are ingoing after the collapse.
(Something like it should happen also in string theory
if, as many suppose, the trans-Planckian reservoir at the horizon is 
also eliminated there.) Note however that the linear model
described here is surely a gross oversimplification. Turning
on the gravitational interactions between the modes and the 
background, one is led to a picture in which the modes
``dissipate" when propagated backwards in time to the
Planckian regime. Hence what really produces the outgoing 
Hawking quantum must be a complicated collective mode
of the interacting vacuum that ``anti-dissipates" as it approaches
the horizon and turns around. Calculations exploring this in 
quantum gravity were carried out in \cite{collective}.

\subsubsection{Lattice time dependence and geometry fluctuations}
\label{latticetimedep}
The microscopic time dependence of the lattice, i.e. the slow 
spreading of the lattice points, plays a critical
role in transforming an ingoing mode with high Killing frequency to 
an outgoing mode with low Killing frequency, and 
in allowing for mixing of positive and negative frequencies
despite the stationarity of the continuum black hole background.
This suggests the conjecture that in 
quantum gravity the underlying quantum fluctuations of the 
geometry that do not share the stationarity of the black hole
background metric might play this role. A step towards understanding
this might be provided by a two or three dimensional version
of the lattice model, in which the density of lattice points remains
fixed but their microscopic positions fluctuate. This is precisely
what happens with an inhomogeneous flow of a real molecular fluid.
The quantum gravity analysis of  \cite{collective} lends some support
to this conjecture, though it is not clear whether the Lorentz violation
seen there is just an artifact of a non-covariant cutoff or a feature
introduced by the global geometry  of the black hole spacetime.

\subsection{Trans-Planckian question in cosmology}
The modes producing the
inflationary perturbation spectrum
({\it cf.} section  \ref{primalpert}) redshift exponentially
from their trans-Planckian
origins. It has been suggested that this might leave 
a visible imprint on the perturbation spectrum, via a
modified high frequency dispersion relation and/or
a modified initial quantum state for the field modes.
As illustrated
by the Hawking process on the lattice 
however, as long as the redshifting
is adiabatic on the timescale of the modes, they would
remain in their ground states.
If the Hubble rate $H$ during inflation were much
less than the Planck mass $M_P$ (or whatever scale the
modified dispersion sets in)
the modes could be treated
in the standard relativistic manner by the
time the perturbation spectrum is determined. 
At most one might expect an
effect of order $H/M_P$, and 
some analyses suggest the effect will be even smaller. 
This all depends on what {\it state} the modes are in.
They cannot be too far from the vacuum, since otherwise 
running them backwards in time they would develop
an exponentially growing 
energy density which would be incompatible with the
inflationary dynamics. Hence it seems the best one can say 
at present is that there may be room for noticable deviations from the 
inflationary predictions, if $H/M_P$ is large enough.
(See for example \cite{TPC} and references therein
for discussions of these issues.)

\section*{Acknowledgements}
I would like to thank the organizers of the CECS 
School on Quantum Gravity both for a stimulating,
enjoyable experience, and for extraordinary patience. 
I'm also grateful to Brendan Foster, Bei-Lok Hu, and Albert Roura 
for helpful corrections, suggestions, and 
questions about these notes.
This work was supported in part by the NSF under grants
PHY-9800967 and PHY-0300710 at the University of Maryland.

\end{document}